\documentclass[12pt,draftclsnofoot,onecolumn]{IEEEtran}

\usepackage{amsmath,graphics,amssymb,epsfig,subfigure,color,cite}
\usepackage{array}
\usepackage{multirow}
\usepackage{enumerate}
\usepackage{algorithmic}
\usepackage{algorithm}
\usepackage{bm}
\usepackage{float}
\usepackage{makecell}

\newtheorem{thm}{Theorem}
\newtheorem{lem}{Lemma}

\makeatletter
\newcommand{\vast}{\bBigg@{4.5}}
\newcommand{\Vast}{\bBigg@{7.5}}
\makeatother

\begin{document}
    \title{Communication-Efficient Federated Learning over MIMO Multiple Access Channels}

	\author{Yo-Seb Jeon, Mohammad Mohammadi Amiri, and Namyoon Lee
		\thanks{Yo-Seb Jeon is with the Department of Electrical Engineering, POSTECH, Pohang, Gyeongbuk 37673, Republic of Korea (e-mail: yoseb.jeon@postech.ac.kr)}
		\thanks {Namyoon Lee is with the School of Electrical Engineering, Korea University, Seoul, Republic of Korea (e-mail: namyoon@korea.ac.kr).}
		\thanks{Mohammad Mohammadi Amiri is with the Media Laboratory, Massachusetts Institute of Technology, Cambridge, MA 02139, USA (e-mail: mamiri@mit.edu).}
	}
	\vspace{-2mm}	
	
	\maketitle
	\vspace{-12mm}

	\begin{abstract} 

Communication efficiency is of importance for wireless federated learning systems. In this paper, we propose a communication-efficient strategy for federated learning over multiple-input multiple-output (MIMO) multiple access channels (MACs). The proposed strategy comprises two components. When sending a locally computed gradient, each device compresses a high dimensional local gradient to multiple lower-dimensional gradient vectors using block sparsification. When receiving a superposition of the compressed local gradients via a MIMO-MAC, a parameter server (PS) performs a joint MIMO detection and the sparse local-gradient recovery. Inspired by the turbo decoding principle, our joint detection-and-recovery algorithm accurately recovers the high-dimensional local gradients by iteratively exchanging their beliefs for MIMO detection and sparse local gradient recovery outputs. We then analyze the reconstruction error of the proposed algorithm and its impact on the convergence rate of federated learning. From simulations, our gradient compression and joint detection-and-recovery methods diminish the communication cost significantly while achieving identical classification accuracy for the case without any compression.
	\end{abstract}

	\begin{IEEEkeywords}
		Federated learning, distributed machine learning, gradient compression, gradient reconstruction, compressed sensing.
	\end{IEEEkeywords}
	
	\section{Introduction}\label{Sec:Intro}
	Federated learning is a distributed machine learning technique for training a global model at a parameter server (PS) through collaboration with wireless devices, each with its own local training data set, where the local data never leaves the devices. \cite{Konecny:15,McMahan:17,Niknam:19,Zhu:20,Gunduz:20}. 
	In federated learning, each wireless device updates a \textit{local} model based on its local training data set and then sends the local model update (e.g., a gradient vector of the local model) to the PS. 
	The PS then updates a \textit{global} model by aggregating the local model updates and then broadcasts the updated global model to the devices. 
	By repeating this collaboration between the PS and the devices, the PS is able to train the global model without directly accessing the devices' data, thereby enhancing the privacy of data generated at the devices.  
	Thanks to this advantage, federated learning has received a great deal of attention as a viable solution for privacy-sensitive machine learning applications at the wireless edge \cite{Niknam:19,Zhu:20,Gunduz:20,Samarakoon:19,Chen:19}. 

    One of the primary research goals in federated learning is to improve communication efficiency in transmitting the local model updates from the devices. 
    The primary reason is that the communication overhead required by these local updates increases with the number of parameters representing the global model, which is typically an enormous value.
    To reduce the communication overhead, gradient compression methods based on lossy compression of the local gradient vectors computed at the devices have been widely considered in the literature \cite{Konecny:16,Alistarh:17,SignSGD,Lee:21,UVeqFed,Du:20,Aji:17,Wangni:18,Lin:18,Amiri:TSP,Amiri:TWC,FedQCS}.  
    Two representative approaches for gradient compression are \textit{gradient quantization} and \textit{gradient sparsification}.
    In gradient quantization, local gradients are quantized and then transmitted using digital transmission \cite{Konecny:16,Alistarh:17,SignSGD,Lee:21,UVeqFed,Du:20}. 
    The most widely adopted technique is one-bit quantization, in which the device only transmits the sign of each local gradient entry. 
    It is shown in \cite{SignSGD} that aggregating the signs of different local gradient vectors based on majority voting achieves the same variance reduction as transmission without quantization. 
    This scalar quantization technique is extended in \cite{UVeqFed,Du:20} to exploit vector quantization.
    In these works, a high-dimensional local gradient vector is partitioned and then quantized using vector quantizers such as lattice quantizers \cite{UVeqFed} and Grassmannian quantizers \cite{Du:20}. 
    Gradient sparsification considers sparsifying the gradient by removing less significant elements. 
    Digital transmission with gradient sparsification is studied in \cite{Aji:17,Wangni:18,Lin:18} by adopting an encoding function to leverage the sparsity of the gradient vector.
    Analog transmission with gradient sparsification is developed in \cite{Amiri:TSP,Amiri:TWC} in which the sparsified gradient vector is compressed by projection onto a lower-dimensional space as in compressed sensing (CS) \cite{Yonina:Book, CS:Book}.
    Gradient compression that exploits the advantages of both gradient quantization and sparsification is also studied in \cite{FedQCS} by leveraging the idea of quantized CS.
    It is reported in \cite{FedQCS} that gradient compression based on quantized CS not only reduces communication overhead for federated learning but also mitigates gradient reconstruction error at the PS. 

  Federated learning over a wireless multiple-access channel (MAC) is another promising solution to reduce communication overhead, particularly when a large number of devices participate in federated learning \cite{Yang:20,Amiri:GlobalSIP,Wen:19,Zhu:TWC,Vu:20,Jeon:21,FL:IRS}. 
    In this solution, multiple devices simultaneously transmit their local model updates (e.g., local gradient vectors) via shared radio resources; thereby, the communication overhead required for the local update transmission does not increase with the number of the devices.
    The most widely adopted approach for federated learning over the wireless MAC is over-the-air computation which exploits the superposition property of the wireless MAC \cite{Yang:20}.
    To further improve the reliability of communication under channel fading, federated learning over a wireless multiple-input multiple-output (MIMO) MAC has recently been studied by developing efficient multi-antenna transmission and/or reception techniques for the PS and/or the wireless devices  \cite{Amiri:GlobalSIP,Wen:19,Zhu:TWC,Vu:20,Jeon:21,FL:IRS}.
    These studies demonstrate that the use of multiple antennas improves model recovery at the PS by mitigating channel fading and interference effects. 
    For example, a multi-antenna technique proposed in \cite{Jeon:21} enables almost perfect reconstruction of the local gradient vectors when the PS is equipped with a massive number of antennas.
    One major challenge, however, is that the communication overhead in the wireless MIMO MAC is still considerable if no gradient compression is applied at the wireless devices. 
    Moreover, despite the potentials of both the gradient compression and simultaneous transmission over the wireless MIMO MAC to reduce communication overhead, none of the existing studies have investigated communication-efficient federated learning to take advantage of both techniques.

    In this work, we study communication-efficient federated learning over a wireless MIMO MAC operating with gradient compression at the wireless devices. 
    We particularly consider the analog gradient compression technique in \cite{Amiri:TSP,Amiri:TWC} in which a local gradient vector is sparsified and then projected onto a lower-dimensional space as in CS \cite{Yonina:Book, CS:Book}. 
    Under this consideration, we reformulate the reconstruction problem of the local gradient vectors from uplink received signals at the PS into a large-scale CS problem with the double-sided linear transformation. To be specific, reconstructing a \textit{sparse} matrix ${\bf G}$ from its noisy linear measurement ${\bf Y} = {\bf H}{\bf G}^{\sf T}{\bf A}^{\sf T} + {\bf Z}$, where ${\bf G}$ is a local gradient matrix whose columns are the local gradient vectors sent from different devices, ${\bf H}$ is a MIMO channel matrix, ${\bf A}$ is a random measurement matrix used for the compression, and ${\bf Z}$ is a channel noise matrix.
    Existing CS algorithms to solve this problem require significant computational complexity since the number of the entries of ${\bf G}$ is proportional to both the number of the devices and the number of global model parameters, which can be on the order of millions in federated learning applications. 
    To overcome this limitation, we propose a computationally-efficient algorithm to solve the gradient reconstruction problem at the PS based on a divide-and-conquer strategy. 
    We then analyze the reconstruction error of the proposed algorithm and its impact on the convergence rate of federated learning. 
    Simulation results demonstrate that the proposed algorithm enables almost lossless reconstruction of the local gradient vectors at the PS while achieving significant performance gain over the existing CS algorithms.
	The major contributions of this paper are summarized as follows:
	\begin{itemize}
	    \item We propose a computationally-efficient algorithm to solve the gradient reconstruction problem at the PS for enabling communication-efficient federated learning over a wireless MIMO MAC.
	    Our key observation is that the gradient reconstruction problem in the form of ${\bf Y} = {\bf H}{\bf G}^{\sf T}{\bf A}^{\sf T} + {\bf Z}$ can be decomposed into parallel small-scale CS recovery problems if a compressed gradient matrix ${\bf X} ={\bf A}{\bf G}$ is explicitly estimated from MIMO detection. 
	    Inspired by this observation, the proposed algorithm first estimates the compressed gradient matrix based on the minimum-mean-square-error (MMSE) MIMO detection, then solves small-scale CS recovery problems in parallel by employing the expectation-maximization generalized-approximate-message-passing (EM-GAMP) algorithm in \cite{EMGAMP}. 
	    A major drawback of our divide-and-conquer strategy is that MIMO detection error propagates into the subsequent CS recovery process.
        To alleviate this drawback, the proposed algorithm also leverages a turbo decoding principle that allows the MIMO detection and the CS recovery processes to iteratively exchange their beliefs on the estimate of the compressed gradient matrix. 
        One prominent feature of the proposed algorithm is that the accuracy of the gradient reconstruction at the PS improves as the iteration continues because, in each iteration, the belief from uplink received signals at the PS is complemented by the sparse property of the local gradient vectors. 

		\item We analyze the reconstruction error of the proposed algorithm and its impact on the convergence rate of federated learning. In this analysis, we model each local gradient vector as an independent and identically distributed (IID) random vector whose distribution is assumed to be known at the PS a priori. Under this consideration, we characterize a mean-square-error (MSE) upper bound in reconstructing a global gradient vector when employing the proposed gradient reconstruction algorithm. Furthermore, our analysis demonstrates that the reconstruction error reduces as the gradient compression ratio at the devices decreases and the orthogonality of the channels from the devices to the PS increases. 
		We also characterize the convergence rate of federated learning operating with a stochastic gradient descent (SGD) algorithm. Our result shows that if the reconstruction error at the PS is below a certain level, federated learning converges to a stationary point of a smooth loss function at the rate of $\mathcal{O}\big(\frac{1}{\sqrt{T}}\big)$, where $T$ is the number of total iterations of the SGD algorithm. 
		The trade-off between communication overhead and the convergence rate of federated learning is also captured in our analysis.

	    \item Using simulations, we evaluate the performance of federated learning over a wireless MIMO MAC for an image classification task using the MNIST dataset \cite{MNIST}. 
	    These simulations compare the classification accuracy and the normalized MSE of the proposed gradient reconstruction algorithm with the existing CS recovery algorithms under the same compression ratio and a similar computational complexity. 
	    Simulation results demonstrate that the proposed algorithm achieves almost the same classification accuracy as perfect reconstruction while superior performance to the existing CS algorithms.
	    It is also shown that the proposed algorithm is more robust to the increase in the gradient compression ratio than other algorithms. 
	    We also investigate the effect of the number of turbo iterations and the sparsification level of the gradient compression on the performance of the proposed algorithm. 
	    Based on these simulation results, we demonstrate that using the proposed algorithm at the PS enables accurate reconstruction of the local gradient vectors while effectively reducing the communication overhead by taking advantage of the gradient compression and simultaneous transmission over the wireless MIMO MAC.

	\end{itemize}


	\subsubsection*{Notation}
	Upper-case and lower-case boldface letters denote matrices and column vectors, respectively.
	$\mathbb{E}[\cdot]$ is the statistical expectation,
	$\mathbb{P}(\cdot)$ is the probability,
	and $(\cdot)^{\sf T}$ is the transpose.
	$|\mathcal{A}|$ is the cardinality of set $\mathcal{A}$.
	$({\bf a})_i$ represents the $i$-th entry of vector ${\bf a}$.
	$\|{\bf a}\|\!=\!\sqrt{{\bf a}^{\sf T}{\bf a}}$ is the Euclidean norm of a real vector ${\bf a}$.
	$\mathcal{N}({\bm \mu},{\bf R})$ represents a multivariate Gaussian distribution with mean vector ${\bm \mu}$ and covariance matrix ${\bf R}$, while $\mathcal{N}({\bf x};{\bm \mu},{\bf R})$ denotes the probability density function of a Gaussian random vector ${\bf x}$ with mean vector ${\bm \mu}$ and covariance matrix ${\bf R}$.
	${\bf 0}_n$ and ${\bf 1}_n$ are $n$-dimensional vectors whose elements are zero and one, respectively.

	\begin{figure*}[t]
		\centering 
		{\epsfig{file=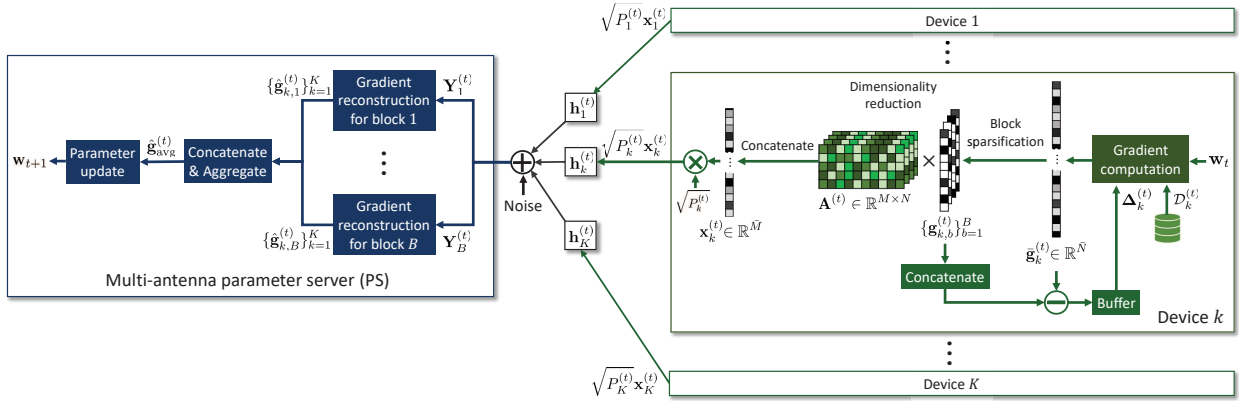, width=16.5cm}}
		\caption{An illustration of uplink communication of federated learning over a wireless MIMO multiple access channel.} \vspace{-7mm}
		\label{fig:System}
	\end{figure*}

	\section{System Model}\label{Sec:System}
	In this section, we present a federated learning framework considered in our work. We then describe uplink transmission and reception procedures in federated learning over a wireless MIMO MAC.

 	\subsection{Federated Learning Framework}	
	We consider federated learning over a wireless MIMO MAC in which a PS equipped with $U$ antennas trains a global model (e.g., deep neural network) by collaborating with $K$ single-antenna wireless devices, as illustrated in Fig.~\ref{fig:System}.
	We assume that the global model is fully represented by a parameter vector ${\bf w}\in \mathbb{R}^{\bar{N}}$, where $\bar{N}$ is the number of the parameters (e.g., the weights and biases of a deep neural network).
    We also assume that the training data samples to optimize the global model are available only at the wireless devices and differ across the devices. 
	Let $\mathcal{D}_k$ be a local training data set available at device $k$, which consists of $|\mathcal{D}_k|$ training data samples.
	A \textit{local} loss function at device $k\in\mathcal{K}=\{1,\ldots,K\}$ for the parameter vector ${\bf w}$ is defined as 
	\begin{align}
	    F_k({\bf w}) = \frac{1}{|\mathcal{D}_k|} \sum_{{\bf u}\in\mathcal{D}_k} f({\bf w};{\bf u}),
	\end{align}
	where $f({\bf w};{\bf u})$ is a loss function computed with a training data sample ${\bf u}\in\mathcal{D}_k$ for the parameter vector ${\bf w}$.
	Then a \textit{global} loss function for the parameter vector ${\bf w}$ is represented as
	\begin{align}\label{eq:global_loss}
	    F({\bf w}) = \frac{1}{\sum_{j=1}^K|\mathcal{D}_j|} 
	    \sum_{{\bf u}\in \cup_j \mathcal{D}_j} f({\bf w};{\bf u})
	    = \frac{1}{\sum_{j=1}^K|\mathcal{D}_j|} \sum_{k=1}^K  |\mathcal{D}_k|  F_k({\bf w}).
	\end{align}

	We focus on a scenario where a gradient-based algorithm (e.g., SGD algorithm) is adopted to optimize the parameter vector for minimizing the global loss function in \eqref{eq:global_loss}.
    We denote $T$ as the number of total iterations of the optimization algorithm, and ${\bf w}_t\in{\mathbb R}^{\bar{N}}$ as the parameter vector at iteration $t$. 
	Each iteration of the optimization algorithm corresponds to one communication round consisting of a pair of downlink and uplink communications. 
    During the downlink communication, the PS broadcasts the current parameter vector to the wireless devices.
    Then, during the uplink communication, the wireless devices transmit the gradient of the parameter vector computed based on their local training data sets to the server.  
    The major focus of our work is to investigate uplink transmission and reception strategies at the server for federated learning; thereby, for simplicity of the analysis, we assume that the downlink communication is error free\footnote{The framework under consideration and the proposed approach can be readily extended to a noisy downlink scenario\cite{Mohammad:DL}}. 
    Under this assumption, all the devices have a globally consistent parameter vector ${\bf w}_t$ for all $t\in\{1,\ldots,T\}$.    
    In what follows, we describe uplink transmission at the wireless devices and uplink reception at the PS during each uplink communication.  

	\subsection{Uplink Transmission at Wireless Devices}	
	For the uplink transmission, each device computes a local gradient vector for the current parameter vector based on its local training data set. 
	A local gradient vector computed at device $k$ is given by
	\begin{align}\label{eq:local_grad}
	    \nabla F_k^{(t)}({\bf w}_t) = \frac{1}{|\mathcal{D}_k^{(t)}|} \sum_{{\bf u}\in\mathcal{D}_k^{(t)}} \nabla f({\bf w}_t;{\bf u}),
	\end{align}
	where $\mathcal{D}_k^{(t)} \subset \mathcal{D}_k$ is a mini-batch randomly drawn from $\mathcal{D}_k$, and $\nabla$ is a gradient operator. 
	We assume that the mini-batch size at device $k$ is fixed as $D_k^{\rm mini}=|\mathcal{D}_k^{(t)}|$  for all $t\in\{1,\ldots,T\}$.
    Since direct transmission of the local gradient vector in \eqref{eq:local_grad} imposes large communication overhead when $\bar{N} \gg 1$, we assume that each device applies \textit{lossy} compression to its local gradient vector before uplink transmission. 
    Details of the gradient compression technique considered in our work will be elaborated in Sec.~\ref{Sec:Comp}. 
    Let ${\bf x}_k^{(t)} = \big[x_{k}^{(t)}[1], \cdots, x_{k}^{(t)}[\bar{M}]\big]^{\sf T} \in \mathbb{R}^{\bar{M}}$ be a \textit{compressed} gradient vector determined at device $k$ with $\bar{M}<\bar{N}$.
    Then an uplink signal vector transmitted from device $k$ at iteration $t$ is denoted by $\tilde{\bf x}_k^{(t)} = \sqrt{P_k^{(t)}}{\bf x}_k^{(t)}$, where $P_k^{(t)}$ is a power scaling factor for device $k$ at iteration $t$. 
    In particular, the power scaling factor is set to be $P_k^{(t)} = \frac{\bar{M}}{\|{\bf x}_k^{(t)}\|^2}$, so that  an average power of transmitting each entry of the uplink signal vector becomes one (i.e., $\mathbb{E}[|(\tilde{\bf x}_{k}^{(t)})_m|^2] = 1$, $\forall m$). 
    All $K$ devices simultaneously transmit their uplink signal vectors over a wireless MAC by utilizing  $\bar{M}$ shared radio resources that are orthogonal in either the time or the frequency domain.
	We assume that the power scaling factor is separately conveyed to the PS in an error-free fashion because the overhead to transmit this scalar value is negligible compared to communication overhead to transmit the compressed gradient vector.

	\subsection{Uplink Reception at Parameter Server} 
    Once the uplink transmission of the wireless devices ends, the PS reconstructs the local gradient vectors from uplink received signals. 
    Let ${\bf x}^{(t)}[m] = \big[x_{1}^{(t)}[m], x_{2}^{(t)}[m],\cdots, x_{K}^{(t)}[m]\big]^{\sf T}$ be the aggregation of the compressed gradient entries sent by $K$ devices at the $m$-th resource, where $x_{k}[m]$ is the $m$-th element of ${\bf x}_k^{(t)}$.
    Then an uplink received vector at the PS for the $m$-th resource is given by
    \begin{align}\label{eq:received_signal}
        {\bf y}^{(t)}[m] ={\bf H}^{(t)}({\bf P}^{(t)})^{1/2} {\bf x}^{(t)}[m] + {\bf z}^{(t)}[m] = \tilde{\bf H}^{(t)}{\bf x}^{(t)}[m] + {\bf z}^{(t)}[m],
	\end{align}
	where ${\bf P}^{(t)} = {\sf diag}\big(P_1^{(t)},\ldots,P_K^{(t)}\big)$ is a power allocation matrix, ${\bf H}^{(t)}\in\mathbb{R}^{U \times K}$ is a MIMO channel matrix from $K$ devices to the server, $\tilde{\bf H}^{(t)} \triangleq {\bf H}^{(t)}({\bf P}^{(t)})^{1/2}$, and ${\bf z}^{(t)}[m] \sim \mathcal{N}({\bf 0}_U , \sigma^2{\bf I}_U)$ is a Gaussian noise vector.
	We assume that the channel matrix remains constant within one communication round (i.e., block-fading assumption) and is perfectly known at the PS via uplink channel estimation. 
	Based on the received signal vectors $\{{\bf y}^{(t)}[m]\}_{m}$, the PS reconstructs the local gradient vectors, $\{{\bf g}_k^{(t)}\}_{k=1}^{K}$, sent from all the $K$ wireless devices. 
	Details of the gradient reconstruction algorithm developed in our work will be elaborated in Sec.~\ref{Sec:Proposed}.

	Let $\hat{\bf g}_k^{(t)}$ be the reconstruction of the local gradient vector ${\bf g}_k^{(t)}$ at the PS.
    Then the PS compute a \textit{global} gradient vector from the reconstructed local gradient vectors as follows:
	\begin{align}\label{eq:global_grad}
	    \hat{\bf g}_{\rm avg}^{(t)} = \sum_{k=1}^K \rho_k \hat{\bf g}_k^{(t)},
	\end{align}
	where $\rho_k \triangleq \frac{D_k^{\rm mini}}{\sum_{j=1}^K D_j^{\rm mini}}$ is assumed to be known at the PS a priori.
	The global gradient vector in \eqref{eq:global_grad} is utilized to update the parameter vector ${\bf w}_t$. 
	For example, if the PS adopts a gradient descent algorithm to optimize the parameter vector, the corresponding update rule is given by  
	\begin{align}\label{eq:global_update}
	    {\bf w}_{t+1} \leftarrow {\bf w}_t - \eta_t \hat{\bf g}_{\rm avg}^{(t)},
	\end{align}
	where $\eta_t >0$ is a learning rate at iteration $t$.     	

	\section{Communication-Efficient Federated Learning over MIMO MAC}
	One major problem of the uplink transmission in federated learning over a wireless MIMO MAC is that direct transmission of the local gradient vector in \eqref{eq:local_grad} imposes considerable communication overhead when the number of the model parameters is large (i.e., $\bar{N} \gg 1$). 
	To alleviate this problem, in this section, we present a gradient compression technique for wireless devices, which takes the advantage of both gradient compression and simultaneous transmission over the wireless MIMO MAC. 
	We then formulate a gradient reconstruction problem at the PS when employing the gradient compression technique at the devices and shed light on the challenges of solving this problem.

    \subsection{Gradient Compression at Wireless Devices}\label{Sec:Comp} 
    The key idea of the gradient compression technique is to sparsify a local gradient vector in a block-wise fashion and then compress the sparsified vector as in CS \cite{Yonina:Book, CS:Book}.
    This technique is a simple variant of the gradient compression technique in \cite{Amiri:TSP,Amiri:TWC,FedQCS}.
    Our compression technique consists of two steps, {\em block sparsification} and {\em dimensionality reduction}, as elaborated below.

    {\bf Block Sparsification:}
    The first step of our compression technique is to divide the local gradient vector into  $B$ sub-vectors with dimension $N = \frac{\bar{N}}{B}$, then sparsify each sub-vector by dropping some gradient entries with small magnitudes. 
    Meanwhile, to compensate for information loss caused by the gradient dropping, the dropped gradient entries are stored at each device and then added to the local gradient vector at the next iteration.
    Under the above strategy, the local gradient vector that needs to be sent by device $k$ at iteration $t$ is determined as
	\begin{align}\label{eq:local_grad2}
	    \bar{\bf g}_{k}^{(t)}  =  \nabla F_k^{(t)}({\bf w}_{t})  + {\bm \Delta}_{k}^{(t)}. 
	\end{align}	
    where ${\bm \Delta}_k^{(t)}$ is a residual gradient vector stored by device $k$ before iteration $t$.
    Then the $b$-th local gradient sub-vector at device $k$ is defined as 
	\begin{align}
        \bar{\bf g}_{k,b}^{(t)}
        &= \big[ \bar{g}_{k,\mathcal{I}_{k,b}(1)}^{(t)},\cdots,\bar{g}_{k,\mathcal{I}_{k,b}(N)}^{(t)} \big]^{\sf T},  
    \end{align}		
    where $\bar{g}_{k,n}^{(t)}$ is the $n$-th entry of $\bar{\bf g}_k^{(t)}$, and $\mathcal{I}_{k,b} \subset \bar{\mathcal{I}}\triangleq \{1,\ldots,\bar{N}\}$ is a set of parameter indices associated with the $b$-th sub-vector at device $k$.  
	We assume that $\mathcal{I}_{k,1}, \ldots, \mathcal{I}_{k,B}$ are mutually exclusive subsets of $\bar{\mathcal{I}}$ such that $\bigcup_{b=1}^B \mathcal{I}_{k,b} = \bar{\mathcal{I}}$ and known a priori at the PS. 
    After constructing $B$ sub-vectors, each local gradient sub-vector is sparsified by setting all but the top-$S$ entries with the largest magnitudes as zeros, where $S$ is a sparsification level such that $S \ll N$. We define $S_{\rm ratio} \triangleq S/N$ as a sparsification ratio.
    As a result, device $k$ attains $B$ local gradient sub-vectors $\{{\bf g}_{k,b}^{(t)}\}$, each of which is an $S$-sparse vector.
    We denote the block sparsification step described above as ${\sf BlockSparse}\big(\bar{\bf g}_k^{(t)}\big)$.
   In the sequel, we will show that increasing the number of the gradient sub-vectors, $B$, reduces the computational complexity of the gradient reconstruction process at the PS. However, if $B$ is set too large, the $SB$ non-zero entries in $\{{\bf g}_{k,b}^{(t)}\}$ can significantly differ from the top-$SB$ entries of $\bar{\bf g}_{k}^{(t)} $; this mismatch may lead to a decrease in the performance of federated learning. Therefore, in practical systems, $B$ can be determined according to a target learning performance and the computing power of the PS.

    After the block sparsification, the residual gradient vector $\bar{\bf g}_{k}^{(t)}$ is updated as
	\begin{align}\label{eq:residual}
	    {\bm \Delta}_{k}^{(t+1)}\leftarrow \bar{\bf g}_k^{(t)} - {\sf Concatenate} \big( \{{\bf g}_{k,b}^{(t)}\}_{b=1}^B\big), 
	\end{align}	
	where ${\sf Concatenate}\big( \{{\bf g}_{k,b}^{(t)}\}_{b=1}^B\big) = [g_{k,1}^{\rm cc},\cdots,g_{k,\bar{N}}^{\rm cc}]^{\sf T} \in \mathbb{R}^{\bar{N}}$ such that
	$g_{k,\mathcal{I}_{k,b}(n)}^{\rm cc} = ({\bf g}_{k,b}^{(t)})_{n}$ for all $b, n$.

    {\bf Dimensionality Reduction:}	
    The second step of our compression technique is to compress each sparse sub-vector by random projection onto a lower dimensional space as in CS \cite{Yonina:Book, CS:Book}.
	Let ${\bf x}_{k,b}^{(t)}$ be a compressed gradient sub-vector generated from ${\bf g}_{k,b}^{(t)}$. 
	Then ${\bf x}_{k,b}^{(t)}$ is determined as
	\begin{align}\label{eq:compress_block}
        {\bf x}_{k,b}^{(t)}  = {\bf A}^{(t)} {\bf g}_{k,b}^{(t)},
    \end{align}	
	where ${\bf A}^{(t)} \in \mathbb{R}^{M \times N}$ is a random measurement matrix with $M < N$ at iteration $t$.
	In this work, we set ${\bf A}^{(t)}$ as an IID random matrix with $({\bf A}^{(t)})_{m,n}\sim \mathcal{N}(0,1/M)$ which provides promising properties (e.g., restricted isometry property, null space property) for guaranteeing accurate recovery of a sparse signal \cite{CS:Book}.
	Furthermore, ${\bf A}^{(t)}$ is randomly and independently drawn for each iteration, in order to avoid a null-space problem caused by a fixed random measurement matrix, as discussed in \cite{RandomA}.
	
	By concatenating these $B$ compressed sub-vectors, a compressed gradient vector for device $k$ at iteration $t$ is finally obtained as 
	\begin{align}\label{eq:compress_total}
        {\bf x}_k^{(t)}
        = \big[({\bf x}_{k,1}^{(t)})^{\sf T}, \cdots, ({\bf x}_{k,B}^{(t)})^{\sf T}\big]^{\sf T} \in\mathbb{R}^{\bar{M}}.
    \end{align}	
    The dimension of the compressed vector is $\bar{M}=BM$; thereby, the compression ratio of our  technique is $R=\frac{N}{M} > 1$.
    The overall procedures of our gradient compression technique is illustrated in Fig.~\ref{fig:System} and also summarized in Steps 5--8 in Algorithm~\ref{alg:Framework}.

    A significant advantage of our gradient compression technique is that it requires $R$ times lower communication overhead than the direct transmission of the local gradient vector, utilizing $\bar{N}$ radio resources per device. 
    Another prominent advantage is that the communication overhead of our technique does not increase with the number of devices participating in federated learning because all the devices utilize the same resources available in the wireless MAC.
    Therefore, our compression technique effectively reduces the communication overhead of federated learning with a large number of devices.

	\begin{algorithm}[ht]
		\caption{Communication-Efficient Federated Learning Framework}\label{alg:Framework}
		{\small
	    \hspace*{\algorithmicindent} \textbf{Input}: ${\bf w}_1 \in \mathbb{R}^{\bar{N}}$, ${\bf A}\in \mathbb{R}^{M\times N}$, $\{\mathcal{I}_{k,b}\}_{k,b}$  \\
        \hspace*{\algorithmicindent} \textbf{Output}: ${\bf w}_T \in \mathbb{R}^{\bar{N}}$		
		{\begin{algorithmic}[1]
            \STATE Initialize ${\bf \Delta}_{k}^{(1)}={\bf 0}_{\bar{N}}$, $\forall k$. 
			\FOR {$t=1$ to $T$}
			    \STATE \!\!\!{\em At the wireless devices:}
				\FOR {Each device $k\in\mathcal{K}$}
    				\STATE $\bar{\bf g}_{k}^{(t)} = \nabla F_{k}^{(t)}\big({\bf w}_t\big)  + {\bf \Delta}_{k}^{(t)}$.
    				\STATE $\{{\bf g}_{k,b}^{(t)}\}_{b=1}^B = {\sf BlockSparse}\big(\bar{\bf g}_{k}^{(t)}\big)$.
                    \STATE ${\bf \Delta}_{k}^{(t+1)} = \bar{\bf g}_{k}^{(t)} - {\sf Concatenate}\big(\{{\bf g}_{k,b}^{(t)}\}_{b=1}^B\big)$.
                    \STATE ${\bf x}_{k,b}^{(t)}={\bf A}^{(t)}{\bf g}_{k,b}^{(t)}$, $\forall b$.
                    \STATE Transmit $\tilde{\bf x}_k^{(t)} = \sqrt{P_k^{(t)}}{\bf x}_k^{(t)}$ to the PS via $\bar{M}$ shared radio resources.
		        \ENDFOR
    			\STATE \!\!\!{\em At the parameter server:}
    			    \STATE Construct ${\bf Y}_b^{(t)}$ from $\big\{{\bf y}^{(t)}[m]\big\}_{m=1}^{\bar{M}}$, $\forall b$.
			        \STATE $\{\hat{\bf g}_{k,b}^{(t)}\}_{k\in\mathcal{K}} = {\sf GradReconst}\big({\bf Y}_b^{(t)}\big)$ from {\bf Algorithm~\ref{alg:Proposed}}, $\forall b$.
			        \STATE $\hat{\bf g}_{k}^{(t)} = {\sf Concatenate}\big(\{\hat{\bf g}_{k,b}^{(t)}\}_{b=1}^B\big)$, $\forall k$.
    			    \STATE $\hat{\bf g}_{\rm avg}^{(t)} = \sum_{k=1}^K \rho_k \hat{\bf g}_{k}^{(t)}$.
    			\STATE ${\bf w}_{t+1} = {\bf w}_t - \eta_t \hat{\bf g}_{\rm avg}^{(t)}$.
    			\STATE Broadcast ${\bf w}_{t+1}$ to the wireless devices.
			\ENDFOR
		\end{algorithmic}}}
	\end{algorithm}

    {\bf Remark 1 (Digital Transmission Strategy):}
    In this work, we focus only on analog transmission at the wireless devices in which the uplink signal vector $\tilde{\bf x}_k^{(t)}$ is transmitted without digital modulation.
    Nevertheless, digital transmission of the uplink signal vector is also feasible by utilizing a digital symbol modulator as a scalar quantizer.
    When employing our gradient compression technique, the distribution of the uplink signal vector can be modeled as $\tilde{\bf x}_k^{(t)} \sim \mathcal{N}({\bf 0}_{\bar{M}},{\bf I}_{\bar{M}})$ for large $N$, which will be justified in Sec.~\ref{Sec:Model}.
    Consequently, a complex-valued signal vector, defined as $\tilde{\bf x}_{{\rm c},k}^{(t)} = \big(\tilde{\bf x}_k^{(t)}\big)_{1:\bar{M}/2} + j\big(\tilde{\bf x}_k^{(t)}\big)_{\bar{M}/2+1:\bar{M}}$, follows $\tilde{\bf x}_{{\rm c},k}^{(t)} \sim \mathcal{CN}({\bf 0}_{\bar{M}/2},0.5{\bf I}_{\bar{M}/2})$ for large $N$.
    This fact implies that each entry of $\tilde{\bf x}_{{\rm c},k}^{(t)}$ can be effectively modulated using quadrature amplitude modulation (QAM).
    Therefore, our gradient compression technique can be compatible with commercial digital communication systems by utilizing the above digital transmission strategy. 

	\subsection{Gradient Reconstruction at Parameter Server}\label{Sec:Problem} 
	When employing the compression technique in Sec.~\ref{Sec:Comp}, the PS faces a new challenge in enabling accurate reconstruction of the global gradient vector in \eqref{eq:global_grad}.
	The underlying reason is that the local gradient vectors are not only compressed, but also simultaneously sent via a wireless MIMO MAC that causes different channel fading across devices.
	To shed light on this challenge, we formulate a gradient reconstruction problem at the PS when employing the compression technique in Sec.~\ref{Sec:Comp}.

    Let $\mathcal{M}_b\triangleq \{(b-1)M+1,\ldots,bM\}$ be a set of resource indices associated with the transmission of the $b$-th compressed sub-vector. 
    Then an uplink received signal matrix associated with the $b$-th compressed sub-vector is given by 
    ${\bf Y}_b^{(t)} = \left[{\bf y}^{(t)}[\mathcal{M}_b(1)],\cdots,{\bf y}^{(t)}[\mathcal{M}_b(M)]\right]$.
    From \eqref{eq:received_signal} and \eqref{eq:compress_total}, the uplink received signal matrix ${\bf Y}_b^{(t)}$ is rewritten as
    \begin{align}\label{eq:received_signal_BL}
	    {\bf Y}_b^{(t)}  
	    = \tilde{\bf H}^{(t)} ({\bf X}_b^{(t)})^{\sf T}  + {\bf Z}_b^{(t)}
	    \overset{(a)}{=} \tilde{\bf H}^{(t)} ({\bf G}_b^{(t)})^{\sf T} ({\bf A}^{(t)})^{\sf T}  + {\bf Z}_b^{(t)},
	\end{align}
	where ${\bf X}_b^{(t)} =\big[ {\bf x}_{1,b}^{(t)} , \cdots, {\bf x}_{K,b}^{(t)}\big]$, 
	${\bf Z}_b^{(t)} = \left[{\bf z}^{(t)}[\mathcal{M}_b(1)],\cdots,{\bf z}^{(t)}[\mathcal{M}_b(M)]\right]$, ${\bf G}_b^{(t)} = \big[ {\bf g}_{1,b}^{(t)}, \cdots, {\bf g}_{K,b}^{(t)}\big]$, and the equality (a) holds because ${\bf X}_b^{(t)} = {\bf A}^{(t)}{\bf G}_b^{(t)}$ from \eqref{eq:compress_block}.
    Note that ${\bf G}_b^{(t)}$ is an $SK$-sparse matrix because ${\bf g}_{1,b}^{(t)} , \ldots, {\bf g}_{K,b}^{(t)}$ are $S$-sparse vectors by the sparsification step during the gradient compression.
	Therefore, the gradient reconstruction problem reduces to $B$ parallel CS recovery problems, each of which aims at finding an $S K$-sparse matrix ${\bf G}_b^{(t)} \in \mathbb{R}^{N \times K}$ from a noisy linear measurement ${\bf Y}_b^{(t)}  \in \mathbb{R}^{U \times M}$.
	The overall procedures of gradient reconstruction and parameter update at the PS are illustrated in Fig.~\ref{fig:System} and also summarized in Steps 12--17 in Algorithm~\ref{alg:Framework}, where ${\sf GradReconst}(\cdot)$ denotes the gradient reconstruction algorithm chosen by the PS.

	\subsection{Limitations of Conventional Approaches}\label{Sec:Limitation}
	The CS recovery problem with the form of \eqref{eq:received_signal_BL} has also been studied for various applications \cite{Peng:05,Yonina:13,Yonina:18}.
    The simplest approach to solve this problem is to formulate an equivalent CS recovery problem by converting the received signal matrix in \eqref{eq:received_signal_BL} into an equivalent vector form:
	\begin{align}\label{eq:received_vector_Kron}
	    {\rm vec}\big({\bf Y}_b^{(t)}\big) 
	    =  {\rm vec}\big(\tilde{\bf H}^{(t)}({\bf G}_b^{(t)})^{\sf T} ({\bf A}^{(t)})^{\sf T}\big)+ {\rm vec}\big({\bf Z}_b^{(t)}\big) 
	    = ({\bf A}^{(t)} \otimes  \tilde{\bf H}^{(t)}){\rm vec}\big(({\bf G}_b^{(t)})^{\sf T}\big)+ {\rm vec}\big({\bf Z}_b^{(t)}\big).
	\end{align}
	Based on this form, an existing compressed sensing algorithm can be applied to recover a sparse vector ${\rm vec}\big(({\bf G}_b^{(t)})^{\sf T}\big) \in\mathbb{R}^{N K}$ from its noisy linear measurement ${\rm vec}\big({\bf Y}_b^{(t)}\big) \in\mathbb{R}^{M U}$.
	This approach, however, requires tremendous computational complexity because the dimensions of both ${\rm vec}\big(({\bf G}_b^{(t)})^{\sf T}\big)$ and ${\rm vec}\big({\bf Y}_b^{(t)}\big)$ are extremely large in many federated learning applications.
    For example, Table~\ref{Table:Complex} in Sec.~\ref{Sec:Simul} shows that the orthogonal matching pursuit (OMP) algorithm to solve the Kronecker-form problem in \eqref{eq:received_vector_Kron} requires the complexity order of $6 \times 10^{15}$ in our simulation with $B=10$, $R=5$, and $S/N=0.04$.
    Another existing approach is to directly solve a matrix-form problem in \eqref{eq:received_signal_BL} by modifying conventional CS algorithms such as  orthogonal matching pursuit (OMP) \cite{Peng:05,Yonina:13}. 
    This approach, however, still requires significant computational complexity that comes from large-scale matrix multiplications.
    For example, Table~\ref{Table:Complex} in Sec.~\ref{Sec:Simul} shows that the OMP algorithm to solve the matrix-form problem in \eqref{eq:received_signal_BL} requires the complexity order of  $4 \times 10^{15}$ in our simulation with $B=10$, $R=5$, and $S/N=0.04$.
    To overcome the limitations of the existing CS approaches, in the following section, we will present a new low-complexity approach to solve the CS recovery problem in \eqref{eq:received_signal_BL}.

	\section{Proposed Gradient Reconstruction Algorithm}\label{Sec:Proposed}
	When employing the gradient compression technique in Sec.~\ref{Sec:Comp}, it is essential to develop a computationally-efficient solution for reconstructing the global gradient vector at the PS while minimizing reconstruction error.
    To achieve this goal, in this section, we propose a novel low-complexity algorithm for solving the gradient reconstruction problem in Sec.~\ref{Sec:Problem}, on the basis of a divide-and-conquer strategy combined with a turbo decoding principle.
    In the rest of this section, we omit the index $t$ for ease of notation.

	\subsection{Basic Principle: Divide-and-Conquer Strategy}\label{Sec:Principle}
	The basic principle of the proposed algorithm is to decompose the gradient reconstruction problem in \eqref{eq:received_signal_BL} into $K$ small-scale CS recovery problems by applying MIMO detection to explicitly estimate a compressed gradient matrix ${\bf X}_b$. 
	To be more specific, the output of the MIMO detection applied to detect ${\bf X}_b$ from a received signal matrix in \eqref{eq:received_signal_BL} can be expressed as
	\begin{align}\label{eq:MIMO_detection}
	    \hat{\bf X}_b = {\bf X}_b + {\bf N}_b,
	\end{align}	
	where  ${\bf N}_{b}$ is a MIMO detection error.
    Then the $k$-th column of $\hat{\bf X}_b$, namely $\hat{\bf x}_{k,b}$, is expressed as 
	\begin{align}\label{eq:small_CS_problem}
	    \hat{\bf x}_{k,b}
	    = {\bf x}_{k,b} + {\bf n}_{k,b}
	    = {\bf A} {\bf g}_{k,b}  + {\bf n}_{k,b},
	\end{align}
	where ${\bf n}_{k,b}$ is the $k$-th column of ${\bf N}_{b}$.
	Since every local gradient sub-vector sent by each device is an $S$-sparse vector as explained in Sec.~\ref{Sec:Comp}, reconstructing ${\bf g}_{k,b}\in\mathbb{R}^N$ from $\hat{\bf x}_{k,b}\in\mathbb{R}^{M}$ is exactly a noisy CS recovery problem which finds an $N$-dimensional sparse vector from an $M$-dimensional observation.  
	Solving this problem in parallel for all $k \in \mathcal{K}$ and $b\in\{1,\ldots,B\}$ requires much smaller complexity than directly solving the CS recovery problem in \eqref{eq:received_signal_BL} which finds a $KN$-dimensional unknown vector from a $UM$-dimensional observation. 
	
\begin{figure*}[t]
		\centering 
		{\epsfig{file=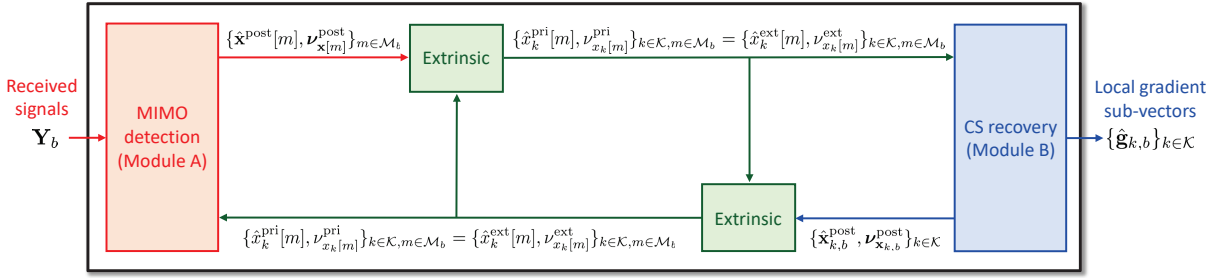, width=16cm}}
		\caption{Block diagram of the proposed gradient reconstruction algorithm applied to block $b$.} \vspace{-7mm}
		\label{fig:Algorithm}
	\end{figure*}

    Despite the advantage in reducing the computational complexity, a major drawback of our strategy is that error in the MIMO detection (e.g., ${\bf N}_b$ in \eqref{eq:MIMO_detection}) propagates into the subsequent CS recovery process. 
    To alleviate this drawback, we leverage a turbo decoding principle that allows the MIMO detection and the CS recovery processes to exchange their beliefs about the entries of the compressed  gradient  matrix.
    The proposed algorithm based on the above principle is illustrated in Fig.~\ref{fig:Algorithm}.

	\begin{figure*}[t]
		\centering 
		\epsfig{file=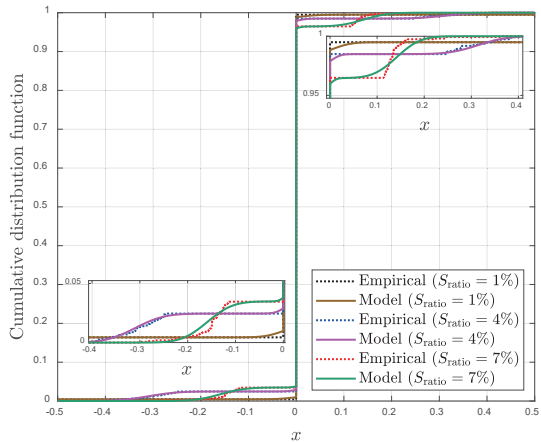, width=7.2cm}
		\caption{Comparison between the empirical cumulative distribution function (CDF) of the local gradient entries sampled from simulation and the CDF of the Bernoulli Gaussian-mixture distribution.} \vspace{-5mm}
		\label{fig:Dist}
	\end{figure*}

	\subsection{Statistical Modeling}\label{Sec:Model}
	To enable the turbo decoding principle based on a Bayesian approach, we establish a statistical model for local gradient sub-vectors as well as for compressed gradient sub-vectors, as done in \cite{FedQCS}. 
	For statistical modeling of the local gradient sub-vectors, we focus on the fact that every local gradient sub-vector is sparse, while its non-zero entries are arbitrarily distributed. 
	We also use the fact that the entries of each local gradient sub-vector are likely to be independent because they are associated with a randomly drawn subset of the model parameters, as explained in Sec.~III-A.
	Motivated by these facts, we model each local gradient sub-vector as an IID random vector whose entry follows the Bernoulli Gaussian-mixture distribution which is well known for its suitability and generality for modeling a sparse random vector \cite{EMGAMP}. 
	Note that the probability density function of the Bernoulli Gaussian-mixture distribution with parameter ${\bm \theta} = \big(\lambda_0,\{\lambda_{\ell},\mu_{\ell},\phi_{\ell}\}_{\ell=1}^L\big)$ is given by 
    \begin{align}\label{eq:GM_model}
        \mathcal{BG}(g; {\bm \theta}) = \lambda_0 \delta(g) + \sum_{\ell=1}^L  \frac{\lambda_{\ell}}{\sqrt{2\pi \phi_{\ell}}} \exp\left(-\frac{(g-\mu_{ \ell})^2}{2\phi_{\ell}}\right),
    \end{align}
    where $\sum_{\ell=0}^L \lambda_{\ell} = 1$.
    As can be seen in \eqref{eq:GM_model}, the above distribution is suitable for modeling the local gradient sub-vector because the sparsity of this sub-vector is effectively captured by the Bernoulli model with parameter $\lambda_0$, while the distribution of non-zero local gradient entries is well approximated by the Gaussian-mixture model with $L$ components.
    How to tune the parameter ${\bm \theta}$ for each gradient sub-vector will be discussed in Sec.~\ref{Sec:EMGAMP}.

    To further justify the Bernoulli Gaussian-mixture model, in Fig.~\ref{fig:Dist}, we compare the {\em empirical} cumulative distribution function (CDF) of the entries of the local gradient sub-vector obtained from simulation\footnote{In this simulation, we employ Algorithm 3 with $R=3$ and $T=30$. Further details of the simulation are described in Sec.~\ref{Sec:Simul}.} with the CDF of the Bernoulli Gaussian-mixture distribution whose parameters are determined by Algorithm 3.
    Fig.~\ref{fig:Dist} shows that the empirical CDF is very similar to the corresponding model-based CDF. This result implies that the local gradient sub-vector can be well modeled by an IID random vector with the Bernoulli Gaussian-mixture distribution.

    On the basis of the Bernoulli Gaussian-mixture model, we also determine a statistical model for the compressed gradient entries. 
    First of all, as can be seen from \eqref{eq:compress_block}, the $m$-th entry of the compressed gradient sub-vector ${\bf x}_{k,b}$ is given by $({\bf x}_{k,b})_m = \sum_{n=1}^N ({\bf A})_{m,n} ({\bf g}_{k,b})_n$.
    This implies that $x_k[m]$ is nothing but the sum of $N$ independent random variables because the random measurement matrix ${\bf A}$ has IID entries, while the local gradient sub-vector ${\bf g}_{k,b}$ can be modeled as an IID random vector.
    Therefore, for large $N$, every compressed gradient entry $x_k[m]$ can be modeled as a Gaussian random variable by the central limit theorem.

	\subsection{Module A: MIMO Detection}\label{Sec:ModuleA}
	In Module A of the proposed algorithm depicted in Fig.~\ref{fig:Algorithm}, the compressed gradient matrix ${\bf X}_b$ is estimated from the uplink received signal matrix ${\bf Y}_b$.
	To minimize the MSE of the estimate of ${\bf X}_b$, we employ the MMSE MIMO detection based on a Gaussian model justified in Sec.~\ref{Sec:Model}.
	The input of Module A is the prior information about the mean and variance of $x_k[m]$, denoted by $\hat{x}_k^{\rm pri}[m] \in \mathbb{R}$ and $\nu_{x_k[m]}^{\rm pri} > 0$, respectively, $\forall k\in \mathcal{K},m \in \mathcal{M}_b$. 
	This information is set approximately at the beginning of the algorithm and is passed from Module B during the turbo iterations of the algorithm.
	As discussed in Sec.~\ref{Sec:Model}, the prior distribution of ${\bf x}[m]$ can be modeled as ${\bf x}[m] \sim \mathcal{N}\big(\hat{\bf x}^{\rm pri}[m], {\bf R}_{{\bf x}[m]}^{\rm pri}\big)$ for large $N$ by the central limit theorem, where $\hat{\bf x}^{\rm pri}[m] = \big[ \hat{x}_1^{\rm pri}[m], \cdots, \hat{x}_K^{\rm pri}[m] \big]^{\sf T}$, ${\bf R}_{{\bf x}[m]}^{\rm pri} = {\sf diag}\big( {\bm \nu}_{{\bf x}[m]}^{\rm pri} \big)$, and ${\bm \nu}_{{\bf x}[m]}^{\rm pri} = \big[ \nu_{x_1[m]}^{\rm pri}, \cdots, \nu_{x_K[m]}^{\rm pri} \big]^{\sf T}$.
	Note that ${\bf R}_{{\bf x}[m]}^{\rm pri}$ is diagonal because $\{x_k[m]\}_{k\in\mathcal{K}}$ are associated with different parameters in the global model.
	Then, the posterior distribution of ${\bf x}[m]$ for a given observation ${\bf y}[m] = \tilde{\bf H} {\bf x}[m] + {\bf v}[m]$ with ${\bf v}[m] \sim \mathcal{N}({\bf 0}_{U},\sigma^2 {\bf I}_U)$ is determined as $p\big({\bf x}[m]\big|{\bf y}[m]\big) = \mathcal{N}({\bf x}[m]; \hat{\bf x}^{\rm post}[m], {\bf R}_{{\bf x}[m]}^{\rm post})$, where the posterior mean and covariance are computed as \cite{Kay:Book}
	\begin{align}
    	\hat{\bf x}^{\rm post}[m] 
    	&= \hat{\bf x}^{\rm pri}[m]  +
    	{\bf R}_{{\bf x}[m]}^{\rm pri}\tilde{\bf H}^{\sf T} {\bm \Omega}_{{\bf x}[m]} \big({\bf y}[m] - \tilde{\bf H} \hat{\bf x}^{\rm pri}[m] \big),  \label{eq:x_hat_post} \\
    	{\bf R}_{{\bf x}[m]}^{\rm post}  
    	&= {\bf R}_{{\bf x}[m]}^{\rm pri} - {\bf R}_{{\bf x}[m]}^{\rm pri}\tilde{\bf H}^{\sf T} {\bm \Omega}_{{\bf x}[m]} \tilde{\bf H}{\bf R}_{{\bf x}[m]}^{\rm pri}, \label{eq:R_x_post}
	\end{align}
	respectively, where ${\bm \Omega}_{{\bf x}[m]}  =  \big( \tilde{\bf H}{\bf R}_{{\bf x}[m]}^{\rm pri}\tilde{\bf H}^{\sf T} + \sigma^2 {\bf I}_{U}\big)^{-1}$. 
	The $k$-th entry of $\hat{\bf x}^{\rm post}[m]$, denoted by $\hat{x}_k^{\rm post}[m]$, is the MMSE estimate of $x_k[m]$ for a given observation ${\bf y}[m]$.
	Meanwhile, the $k$-th diagonal entry of ${\bf R}_{{\bf x}[m]}^{\rm post}$, denoted by $\nu_{x_k[m]}^{\rm post}$, represents the MSE between $x_k[m]$ and $\hat{x}_k^{\rm post}[m]$.

	After computing the MMSE estimates of the compressed gradient entries, Module~A passes its knowledge about these values to Module~B.
	As illustrated in Fig.~\ref{fig:Algorithm}, the prior information utilized in Module~A has been passed from Module~B; thereby, Module~A extracts the \textit{extrinsic} information by excluding the contribution of the prior information from the posterior information. 
	In particular, the extrinsic distribution of  $x_k[m]$ can be determined by assuming that its prior and posterior distributions are given by $p\big(x_k[m]\big) = \mathcal{N}(x_k[m]; \hat{x}_k^{\rm pri}[m],\nu_{x_k[m]}^{\rm pri})$ and $p\big(x_k[m]\big|{\bf y}[m]\big) = \mathcal{N}(x_k[m];\hat{x}_k^{\rm post}[m],\nu_{x_k[m]}^{\rm post})$, respectively.
	Then, by Bayes' rule, the extrinsic distribution of $x_k[m]$ is given by $p\big({\bf y}[m]\big|x_k[m]\big) = \mathcal{N}(x_k[m]; \hat{x}_k^{\rm ext}[m],\nu_{x_k[m]}^{\rm ext})$, where the extrinsic mean and variance are computed as \cite{Guo:11,Liu:19}
	\begin{align}
	    \hat{x}_k^{\rm ext}[m] = \frac{\hat{x}_k^{\rm post}[m]\nu_{x_k[m]}^{\rm pri}  - \hat{x}_k^{\rm pri}[m] \nu_{x_k[m]}^{\rm post}}
        {  \nu_{x_k[m]}^{\rm pri} - \nu_{x_k[m]}^{\rm post}  },~~\text{and}~~
	    \nu_{x_k[m]}^{\rm ext} = \frac{\nu_{x_k[m]}^{\rm pri} \nu_{x_k[m]}^{\rm post}  }
	    { \nu_{x_k[m]}^{\rm pri} - \nu_{x_k[m]}^{\rm post} },   \label{eq:x_hat_nu_ext} 
	\end{align}
	respectively.
    Then the values of $\{\hat{x}_k^{\rm ext}[m], \nu_{x_k[m]}^{\rm ext}\}_{k,m}$ are passed to Module~B which utilizes these values as the prior information about $\{x_k[m]\}_{k,m}$.
    
	\subsection{Module B: CS Recovery}\label{Sec:EMGAMP}
	In Module B of the proposed algorithm, each local gradient sub-vector is reconstructed from the prior information about the compressed gradient entries. 
	To this end, the extrinsic mean and variance of $x_k[m]$ passed from Module A are harnessed as the prior mean and variance of $x_k[m]$, respectively, i.e., 
	$x_k[m]\sim \mathcal{N}(\hat{x}_k^{\rm pri}[m],\nu_{x_k[m]}^{\rm pri})$, where 
    $\hat{x}_k^{\rm pri}[m] = \hat{x}_k^{\rm ext}[m]$ and $\nu_{x_k[m]}^{\rm pri}=\nu_{x_k[m]}^{\rm ext}$, $\forall k \in \mathcal{K},m\in\mathcal{M}_b$.
    Then the prior mean and covariance of ${\bf x}_{k,b}$ are determined as $\hat{\bf x}_{k,b}^{\rm pri} = \big[\hat{x}_{k}^{\rm pri}[\mathcal{M}_b(1)], \cdots,\hat{x}_{k}^{\rm pri}[\mathcal{M}_b(M)]\big]^{\sf T}$ and ${\bf R}_{{\bf x}_{k,b}}^{\rm pri}	= {\sf diag}\big( {\bm \nu}_{{\bf n}_{k,b}}^{\rm pri} \big)$, respectively, where ${\bm \nu}_{{\bf n}_{k,b}}^{\rm pri} = \big[ {\nu}_{x_k[\mathcal{M}_b(1)]}^{\rm pri}, \cdots,{\nu}_{x_k[\mathcal{M}_b(M)]}^{\rm pri} \big]^{\sf T}$.
    Note that ${\bf R}_{{\bf x}_{k,b}}^{\rm pri}$ is diagonal because $x_k[m_1]$ and $x_k[m_2]$ are associated with different parameters in the global model for $m_1 \neq m_2 \in\mathcal{M}_b$.
    For large $N$ with the IID random measurement matrix ${\bf A}$ in \eqref{eq:compress_block}, all the entries of ${\bf x}_{k,b}$ are statistically identical by the central limit theorem. 
    Motivated by this fact, we approximate the prior covariance as ${\bf R}_{{\bf x}_{k,b}}^{\rm pri} \approx \bar{\nu}_{{\bf x}_{k,b}}^{\rm pri} {\bf I}_M$, where $ \bar{\nu}_{{\bf x}_{k,b}}^{\rm pri}\triangleq \frac{1}{M} \sum_{m\in\mathcal{M}_b} {\nu}_{x_k[m]}^{\rm pri}$.
    Then, with some tedious manipulations from \eqref{eq:x_hat_post}--\eqref{eq:x_hat_nu_ext},
    the prior mean $\hat{\bf x}_{k,b}^{\rm pri}$ is rewritten as an additive white Gaussian noise (AWGN) observation of ${\bf x}_{k,b}$, i.e.,
    \begin{align}\label{eq:ModuleB_eq}
        \hat{\bf x}_{k,b}^{\rm pri} 
        = {\bf x}_{k,b} + {\bf n}_{k,b}^{\rm pri}
        = {\bf A} {\bf g}_{k,b} + {\bf n}_{k,b}^{\rm pri},
    \end{align}	 
    where ${\bf n}_{k,b}^{\rm pri}\sim \mathcal{N}\big({\bf 0}_M,\bar{\nu}_{{\bf x}_{k,b}}^{\rm pri} {\bf I}_M \big)$ is an effective noise independent of ${\bf x}_{k,b}$, as also derived in \cite{Liu:19}.

    To reconstruct the local gradient sub-vector ${\bf g}_{k,b}$ from the prior mean $\hat{\bf x}_{k,b}^{\rm pri}$ in \eqref{eq:ModuleB_eq}, we need to solve a noisy CS recovery problem. 
    Unfortunately, finding the exact MMSE solution to this problem is infeasible as the exact distribution of the local gradient sub-vector is unknown at the PS.
    To circumvent this challenge, we employ the EM-GAMP algorithm in \cite{EMGAMP} based on the statistical model in Sec.~\ref{Sec:Model}.
    This algorithm iteratively computes an approximate MMSE solution to the CS recovery problem by modeling the distribution of an unknown sparse vector using the Bernoulli Gaussian-mixture model; the parameters of this model are also updated iteratively based on the EM principle.
    As discussed in Sec.~\ref{Sec:Model}, the local gradient sub-vectors are well modeled by the Bernoulli Gaussian-mixture distribution; thereby, the EM-GAMP algorithm is a powerful means of solving our CS recovery problem in \eqref{eq:ModuleB_eq}.
    This algorithm also computes the posterior mean and variance of each compressed gradient entry as a byproduct; thereby, it can be directly combined with the turbo decoding principle in the proposed algorithm. 
    In Algorithm~\ref{alg:EMGAMP}, we summarize the EM-GAMP algorithm employed in Module B of the proposed algorithm, where we omit the indices $k$ and $b$ for ease of notation.

    The major steps in Algorithm~\ref{alg:EMGAMP} are described below.
    In Step~3, the pseudo-prior mean and variance of $x_m$ are determined.
    In Step~4, the posterior mean and variance of $x_m$ are computed based on \eqref{eq:ModuleB_eq} and $x_m\sim \mathcal{N}(\hat{p}_m,{\nu}_{p_m})$.
    In Step~6, the estimate of $g_n$ and the corresponding error variance are determined.
    In Step~7, the posterior mean and variance of $g_n$ are computed under the assumptions of $g_n\sim \mathcal{BG}(g;{\bm \theta})$ and $\hat{r}_n = g_n + \xi_n$ with $\xi_n\sim \mathcal{N}(0,{\nu}_{r_n})$ as derived in \cite{EMGAMP}, where $\beta_{n,0} = \lambda_0 \mathcal{N}(0; \hat{r}_n,\nu_{r_n}) $,  $\beta_{n,\ell} =  \lambda_{\ell} \mathcal{N}( \hat{r}_n;  \mu_{\ell},\nu_{r_n} + \phi_{\ell})$, 
    \begin{align*}
        \lambda_{n,0}^\prime &= \frac{\beta_{n,0}}{\beta_{n,0} + \sum_{i=1}^L \beta_{n,i}},~~
        \lambda_{n,\ell}^\prime = \frac{\beta_{n,\ell}}{\beta_{n,0} + \sum_{i=1}^L \beta_{n,i}},~~ 
        {\mu}_{n,\ell}^\prime &= \frac{\hat{r}_n\phi_{\ell}   + \mu_{\ell}\nu_{r_n} } 
            {\nu_{r_n}  + \phi_{\ell}},~~
         {\phi}_{n,\ell}^\prime = \frac{\nu_{r_n}\phi_{\ell}}  {\nu_{r_n}  + \phi_{\ell}},   
    \end{align*} 
    for all $\ell\in\{1,\ldots,L\}$.
    In Step~8, the parameters of the Bernoulli Gaussian-mixture model are computed based on the EM principle as derived in \cite{EMGAMP}, where $\lambda_0'' = \frac{1}{N}\sum_{n=1}^{N} \lambda_{n,0}^\prime$,
    \begin{align}\label{eq:EM_params}
        \lambda_{\ell}'' \approx
         \frac{1}{N}\sum_{n=1}^{N} \lambda_{n,\ell}^\prime  ,~~
         \mu_{\ell}'' \approx
        \frac{ \sum_{n=1}^{N} \lambda_{n,\ell}^\prime \mu_{n,\ell}^\prime } 
        { \sum_{n=1}^{N} \lambda_{n,\ell}^\prime},~~
        \phi_{\ell}'' \approx
        \frac{ \sum_{n=1}^{N} \lambda_{n,\ell}^\prime \left( |\mu_{\ell} - \mu_{n,\ell}^\prime |^2 + \phi_{n,\ell}^\prime \right) } 
        { \sum_{n=1}^{N} \lambda_{n,\ell}^\prime}.
    \end{align}	
    The asymptotic convergence of the EM-GAMP algorithm was studied in \cite{EMGAMP2}, in which the authors showed that under certain conditions, the performance of this algorithm asymptotically coincides with that of the original GAMP algorithm with perfect knowledge of the input distribution. We refer interested readers to \cite{EMGAMP2} for more details. Meanwhile, the complexity order of the EM-GAMP algorithm is given by $\mathcal{O}(MN)$ in terms of the number of real multiplications (see Step 3 and Step 6 of Algorithm~\label{alg:EMGAMP}).

\begin{algorithm}[ht]
	\caption{EM-GAMP Algorithm}\label{alg:EMGAMP}
	{\small 
	\hspace*{\algorithmicindent} ~\textbf{function} ${\sf EMGAMP}\big(\hat{\bf g}, {\bm \nu}_{{\bf g}}, {\bm \theta}=(\lambda_0,\{\lambda_{\ell},\mu_{\ell},\phi_{\ell}\}_{\ell=1}^L),\hat{\bf x}^{\rm pri}, \bar{\nu}_{{\bf x}}^{\rm pri} \big)$ 
	\begin{algorithmic}[1]
    \STATE Set $\hat{s}_m = 0$, $\hat{g}_n^{\rm old} = \hat{g}_n$, and $A_{m,n} = ({\bf A})_{m,n}$, $\forall m,n$.
	\FOR {$i=1$ to $I_{\rm GAMP}$} 
        \STATE Set $\hat{p}_m = \sum_{n=1}^{N} A_{m,n} \hat{g}_n - \nu_{p_m} \hat{s}_m$ and $\nu_{p_m} = \sum_{n=1}^{N} |A_{m,n}|^2 \nu_{g_n}$, $\forall m$.
        \STATE Compute $\hat{x}^{\rm post}[m] = ({\hat{p}_m \bar{\nu}_{\bf x}^{\rm pri}  + \hat{x}^{\rm pri}[m] \nu_{p_m}  })/({\nu_{p_m} + \bar{\nu}_{\bf x}^{\rm pri}})$ and $\nu_{x[m]}^{\rm post} =\big(1/{ \nu_{p_m} + 1/\bar{\nu}_{\bf x}^{\rm pri}}\big)^{-1}$, $\forall m$.
        \STATE Set $\hat{s}_m = (\hat{x}^{\rm post}[m] - \hat{p}_m)/\nu_{p_m}$ and $\nu_{s_m} = (1-\nu_{x[m]}^{\rm post}/\nu_{p_m}) / \nu_{p_m}$, $\forall m$.
        \STATE Set $\hat{r}_n = \hat{g}_n + \nu_{r_n} \sum_{m=1}^{M} A_{m,n} \hat{s}_m$ and $\nu_{r_n} = \big( \sum_{m=1}^{M} |A_{m,n}|^2 \nu_{s_m}\big)^{-1}$, $\forall n$.
        \STATE Compute $\hat{g}_n = \sum_{\ell=1}^L \lambda_{n,\ell}^\prime {\mu}_{n,\ell}^\prime$ and $\nu_{g_n} =  \sum_{\ell=1}^L \lambda_{n,\ell}^\prime  \big( {\phi}_{n,\ell}^\prime + ({\mu}_{n,\ell}^\prime)^2\big) - (\hat{g}_n)^2$, $\forall n$.
        \STATE Update ${\bm \theta} \leftarrow (\lambda_0'',\{\lambda_{\ell}'',\mu_{\ell}'',\phi_{\ell}''\}_{\ell=1}^L)$ from \eqref{eq:EM_params}.
        \STATE Break if $\sum_{n=1}^{N} (\hat{g}_n^{\rm old} - \hat{g}_n)^2  < \tau_{\rm GAMP} \sum_{n=1}^{N} (\hat{g}_n^{\rm old} )^2$.
        \STATE Set $\hat{g}_n^{\rm old} = \hat{g}_n$, $\forall n$.
	\ENDFOR
	\STATE Set $\hat{\bf g} = [\hat{g}_1,\cdots,\hat{g}_{N}]^{\sf T}$ and
	    ${\bm \nu}_{{\bf g}} = [\nu_{g_1},\cdots,\nu_{g_N} ]^{\sf T}$.
    \STATE Set
	    $\hat{\bf x}^{\rm post} = [\hat{x}^{\rm post}[1],\cdots,\hat{x}^{\rm post}[M]]^{\sf T}$ and  
	    ${\bm \nu}_{\bf x}^{\rm post} = [\nu_{x[1]}^{\rm post} ,\cdots,\nu_{x[M]}^{\rm post} ]^{\sf T}$.
    \STATE {\bf return} $\hat{\bf g}, {\bm \nu}_{{\bf g}}, {\bm \theta},\hat{\bf x}^{\rm post}, {\bm \nu}_{{\bf x}}^{\rm post}$.
\end{algorithmic}}
\end{algorithm}        

Once the EM-GAMP algorithm applied to $\hat{\bf x}_{k,b}^{\rm pri}$ converges, Module~B acquires not only the approximate MMSE estimate of $\hat{\bf g}_{k,b}$, but also the posterior mean and variance of compressed gradient entries, given by $\{\hat{x}_{k}^{\rm post}[m],{\nu}_{x_k[m]}^{\rm post}\}_{m\in\mathcal{M}_b}$ (see Step~4 in Algorithm~\ref{alg:EMGAMP}). 
As done in Module A, the extrinsic mean and variance, denoted by $\hat{x}_{k}^{\rm ext}[m]$ and ${\nu}_{x_k[m]}^{\rm ext}$, respectively, are computed from $\hat{x}_{k}^{\rm post}[m]$ and ${\nu}_{x_k[m]}^{\rm post}$ according to \eqref{eq:x_hat_nu_ext}, $\forall k \in \mathcal{K},m\in\mathcal{M}_b$.
Then these values are passed to Module A which utilizes these values as the prior information about $\{x_k[m]\}_{k,m}$. 

\begin{algorithm}[ht]
	\caption{Proposed Gradient Reconstruction Algorithm}\label{alg:Proposed}
	{\small 
	\hspace*{\algorithmicindent} ~\textbf{function} ${\sf GradReconst}\big( {\bf Y}_b \big)$ 
	\begin{algorithmic}[1]
	\STATE Initialize $\hat{\bf g}_{k,b}, {\bm \nu}_{{\bf g}_{k,b}}, {\bm \theta}_{k,b}, \hat{x}_k^{\rm pri}[m], \nu_{x_k[m]}^{\rm pri}$,~$\forall k,m \in\mathcal{M}_b$. 
	\FOR {$i=1$ to $I_{\rm Turbo}$}
	    \FOR {$m \in \mathcal{M}_b$}
	        \STATE Set $\hat{\bf x}^{\rm pri}[m] = \big[\hat{x}_1^{\rm  pri}[m],\cdots,\hat{x}_K^{\rm  pri}[m]\big]^{\sf T}$ 
    	        and ${\bf R}_{{\bf x}[m]}^{\rm pri} = {\sf diag}\big( \big[\nu_{x_1[m]}^{\rm pri} ,\ldots,\nu_{x_K[m]}^{\rm pri} \big]\big)$.
    	    \STATE Compute $\hat{\bf x}^{\rm post}[m]$ from \eqref{eq:x_hat_post} and ${\bf R}_{{\bf x}[m]}^{\rm post}$ from \eqref{eq:R_x_post}.
	    \ENDFOR
	    \STATE Update $\hat{x}_k^{\rm pri}[m] \leftarrow \hat{x}_k^{\rm ext}[m]$ and $\nu_{x_k[m]}^{\rm pri} \leftarrow \nu_{x_k[m]}^{\rm ext}$ from \eqref{eq:x_hat_nu_ext},~$\forall k,m  \in \mathcal{M}_b$.
	    \FOR {$k=1$ to $K$}
	            \STATE Set $\hat{\bf x}_{k,b}^{\rm pri} = \big[\hat{x}_k^{\rm  pri}[(b-1)M +1],\cdots,\hat{x}_k^{\rm  pri}[bM]\big]^{\sf T}$ 
    	        and $\bar{\nu}_{{\bf x}_{k,b}}^{\rm pri} = \frac{1}{M}\sum_{m\in \mathcal{M}_b}\nu_{x_k[m]}^{\rm pri}$.
    	        \STATE $\big\{\hat{\bf g}_{k,b}, {\bm \nu}_{{\bf g}_{k,b}}, {\bm \theta}_{k,b},\hat{\bf x}_{k,b}^{\rm post}, {\bm \nu}_{{\bf x}_{k,b}}^{\rm post}\big\} = {\sf EMGAMP}(\hat{\bf g}_{k,b}, {\bm \nu}_{{\bf g}_{k,b}}, {\bm \theta}_{k,b},\hat{\bf x}_{k,b}^{\rm pri}, \bar{\nu}_{{\bf x}_{k,b}}^{\rm pri})$.
	    \ENDFOR
	    \STATE Update $\nu_{x_k[m]}^{\rm pri} \leftarrow \nu_{x_k[m]}^{\rm ext}$
	    and $\hat{x}_k^{\rm pri}[m] \leftarrow \hat{x}_k^{\rm ext}[m]$ from \eqref{eq:x_hat_nu_ext},~$\forall k,m \in \mathcal{M}_b$. 
	\ENDFOR
	\STATE 	\textbf{return} $\hat{\bf g}_{1,b},\hat{\bf g}_{2,b},\ldots,\hat{\bf g}_{K,b}$.
\end{algorithmic}}
\end{algorithm}

	\subsection{Summary}
	In Algorithm~\ref{alg:Proposed}, we summarize the operations of the proposed algorithm that iterates between Module A and Module B based on the turbo decoding principle. 
	We denote $I_{\rm Turbo}$ as the total number of the iterations of the proposed algorithm.
    As can be seen in Algorithm~\ref{alg:Proposed}, in Module~A, the compressed gradient entries are estimated based on the observation of the uplink received signals. Meanwhile, in Module B, the estimates of the compressed gradient entries are refined by exploiting their sparse property. Since Module A and Module B utilize different sources of information for estimating the compressed gradient entries, these modules can complement each other by iteratively exchanging their extrinsic beliefs about the compressed gradient entries.
    Recall that the reduction in the estimation error of the compressed gradient entries leads to the decrease in the noise in the CS recovery process for reconstructing the local gradient sub-vectors, as discussed in Sec.~\ref{Sec:Principle}. 
    Therefore, the use of the turbo decoding principle in the proposed algorithm improves the reconstruction accuracy of the local gradient sub-vectors at the PS.



	\section{Performance Analysis}\label{Sec:Analysis}
    In this section, we analyze the reconstruction error of the proposed algorithm and its impact on the convergence rate of federated learning over a wireless MIMO MAC. 

	\subsection{Reconstruction Error Analysis}\label{Sec:Error}
	In this analysis, we characterize the MSE of a global gradient vector reconstructed from the proposed algorithm. 
	To this end, we model each local gradient sub-vector as an IID random vector, which can be justified using the Bernoulli Gaussian-mixture distribution, as discussed in Sec.~\ref{Sec:Model}.
	Meanwhile, we assume that the distribution of the local gradient sub-vectors can differ across devices and blocks, while the mean and variance of each gradient sub-vector are known\footnote{To realize this assumption, the information of the means and variances of the local gradient sub-vectors needs to be conveyed from the devices to the PS. Fortunately, the overhead to transmit this information is negligible compared to communication overhead to transmit the compressed gradient vectors.} at the PS.
    Under these assumptions, extrinsic means and variances passed from Module A of the proposed algorithm are characterized as given in the following lemma:
    
    \vspace{1mm}
    \begin{lem}\label{Lem1}
        Let $\hat{x}_k^{(t)}[m]$ be the extrinsic mean of $x_k^{(t)}[m]$ computed from Module A of the proposed algorithm for a given observation ${\bf y}^{(t)}[m]$.
        For large $N$, an extrinsic mean vector $\hat{\bf x}_{k,b}^{(t)} = \big[\hat{x}_k^{(t)}[\mathcal{M}_b(1)],\cdots, \hat{x}_k^{(t)}[\mathcal{M}_b(M)] \big]^{\sf T}$ is an AWGN observation of ${\bf x}_{k,b}^{(t)}$ with noise variance $\nu_{k,b}^{(t)}$, i.e., 
        \begin{align}\label{eq:Lem1}
    	    \hat{\bf x}_{k,b}^{(t)} 
    	    = {\bf x}_{k,b}^{(t)} + {\bf n}_{k,b}^{(t)}~~\text{with}~~{\bf n}_{k,b}^{(t)} \sim\mathcal{N}({\bf 0}_M, \nu_{k,b}^{(t)}{\bf I}_M),
    	\end{align}
		where 
		\begin{align}\label{eq:nu_express_lem1}
            \nu_{k,b}^{(t)}
            &= \frac{1}{{\sf SNR}_k^{(t)}}  \Bigg( \|{\bf h}_k^{(t)}\|^2 -  \sum_{i=1}^{E_{k,b}^{(t)}} \big|\big({\bf u}_{k,b,i}^{(t)}\big)^{\sf T} {\bf h}_k^{(t)}\big|^2  \frac{\lambda_{k,b,i}^{(t)} }{\lambda_{k,b,i}^{(t)} + \sigma^2}  \Bigg)^{-1},
        \end{align}
        ${\sf SNR}_k^{(t)} = \frac{P_k}{\sigma^2}$,  $\lambda_{k,b,i}^{(t)}$ is the $i$-th largest eigenvector of ${\bm \Phi}_{k,b}^{(t)}$, ${\bf u}_{k,b,i}^{(t)}$ is the eigenvector associated with $\lambda_{k,b,i}^{(t)}$, $E_{k,b}^{(t)}$ is the rank of ${\bm \Phi}_{k,b}^{(t)}$,
        ${\bm \Phi}_{k,b}^{(t)} = R\sum_{j\neq k} \zeta_{j,b}^{(t)} P_j^{(t)}{\bf h}_j^{(t)}\big({\bf h}_j^{(t)}\big)^{\sf T}$,
    	${\bf h}_k^{(t)}$ is the $k$-th column of ${\bf H}^{(t)}$, and $\zeta_{j,b}^{(t)}$ is the variance of each entry of ${\bf g}_{j,b}^{(t)}$. 
    \end{lem}
    \begin{IEEEproof}
         See Appendix~\ref{Apdx:Lem1}.
    \end{IEEEproof}
    \vspace{1mm}

    Utilizing the result in Lemma~\ref{Lem1}, we characterize an MSE upper bound for the global gradient vector reconstructed by the proposed algorithm.
    This result is given in the following theorem:
    
    \vspace{1mm}
    \begin{thm}\label{Thm1}
        In the asymptotic regime of $N \rightarrow \infty$ and $N/M \rightarrow R$ for some $R \geq 1$, the global gradient vector, $\hat{\bf g}_{\rm avg}^{(t)}$, reconstructed by the proposed algorithm with $I_{\rm Turbo}=1$ satisfies the following MSE bound: 
        \begin{align}\label{eq:Thm1}
        \mathbb{E}\big[\| {\bf g}_{\rm avg}^{(t)} - \hat{\bf g}_{\rm avg}^{(t)}\|^2\big]
        \leq   \sum_{k=1}^K \sum_{b=1}^B  \rho_k^2 
        N \zeta_{k,b}^{(t)} \left(1 - \frac{\zeta_{k,b}^{(t)}}{R\zeta_{k,b}^{(t)} + \nu_{k,b}^{(t)} } \right), 
        \end{align}
        where ${\bf g}_{\rm avg}^{(t)}=\sum_{k=1}^K \rho_k {\bf g}_{k}^{(t)}$ is a global gradient vector at iteration $t$.
    \end{thm}
    \begin{IEEEproof}
        See Appendix~\ref{Apdx:Thm1}.
    \end{IEEEproof}
    \vspace{1mm}

    Theorem~1 shows that the reconstruction error of the proposed algorithm reduces as the compression ratio $R$ approaches to one (i.e., no compression).
    Since the communication overhead increases as the compression ratio decreases, Theorem~1 clearly reveals the trade-off between communication overhead at the devices and reconstruction accuracy at the PS in federated learning over a wireless MIMO MAC.
    Another important observation is that $\nu_{k,b}^{(t)}$ in \eqref{eq:Thm1} decreases not only as the signal-to-noise ratio (SNR) of device $k$ increases, but also as $\big|\big({\bf u}_{k,b,i}^{(t)}\big)^{\sf T} {\bf h}_k^{(t)}\big|^2$ decreases, which can be seen from \eqref{eq:nu_express_lem1}. 
    This observation implies that $\nu_{k,b}^{(t)}$ decreases with the orthogonality between the channel of device $k$ and other devices' channels because 
    $\big|\big({\bf u}_{k,b,i}^{(t)}\big)^{\sf T} {\bf h}_k^{(t)}\big|^2 \rightarrow 0$, $\forall i\in\{1,\ldots,D\}$, as $\big|\big({\bf h}_j^{(t)}\big)^{\sf T} {\bf h}_k^{(t)}\big|^2 \rightarrow 0$, $\forall j\neq k$.
    Therefore, our analytical result demonstrates that the reconstruction error of the proposed algorithm depends on both the gradient compression ratio and  wireless environment (e.g., SNR or channel distribution).
	\subsection{Convergence Rate Analysis}    
    Now, we characterize the convergence rate of federated learning over the wireless MIMO MAC when employing a SGD algorithm. 
    To this end, we make the following assumptions: 
    
    \vspace{1mm}
    {\bf Assumption 1:} The loss function $F({\bf w})$ is $\beta$-smooth and lower bounded by some constant $F({\bf w}^{\star})$, i.e.,  $F({\bf w})\geq F({\bf w}^{\star})$, $\forall {\bf w}\in \mathbb{R}^{\bar{N}}$. 

    {\bf Assumption 2:} 
    For a given parameter vector ${\bf w}_t$, the global gradient vector in \eqref{eq:global_grad} is unbiased, i.e., $\mathbb{E} \big[{\bf g}_{\rm avg}^{(t)} \big| {\bf w}_t \big] = \nabla F({\bf w}_t)$, while its variance is upper bounded as  
    $\mathbb{E} \big[\|{\bf g}_{\rm avg}^{(t)} - \nabla F({\bf w}_t)\|^2 \big| {\bf w}_t \big] \leq \psi  \|\nabla F({\bf w}_t) \|^2$ with a constant $\psi > 0$, for all $t\in\{1,\ldots,T\}$. 
    

    {\bf Rationale:} 
    Assumptions 1--2 are typical for analyzing the convergence rate of the SGD algorithm for a smooth loss function, as considered in \cite{SignSGD,FedQCS}.
    In particular, a variance scaling factor $\psi$ depends on both the mini-batch size and the sparsification level $S$ employed at each device, while this factor decreases with the mini-batch size and the sparsification level. 
    \vspace{1mm}

    Under Assumptions 1--2, we characterize the convergence rate of federated learning, as given in the following theorem:

    \vspace{1mm}
    \begin{thm}\label{Thm2}
        Let $\epsilon_{\rm max} = \max_t \frac{\| {\bf g}_{\rm avg}^{(t)} - \hat{\bf g}_{\rm avg}^{(t)} \|^2}{\| {\bf g}_{\rm avg}^{(t)} \|^2}$ be the maximum normalized reconstruction error of the global gradient vector at the PS. 
        Under Assumptions 1--2, if $\epsilon_{\rm max} < \frac{1}{1+\psi}$, federated learning based on SGD with a learning rate $\eta_t = \frac{\eta}{\sqrt{t}}$ satisfies the following bound:
        \begin{align}\label{eq:Thm2}
            \mathbb{E}\left[ \frac{1}{T} \sum_{t=1}^{T}  \|\nabla F({\bf w}_t)\|^2 \right] \leq 
            \frac{1}{\sqrt{T}}\left[ \frac{F({\bf w}_{1}) - F({\bf w}^\star)}{ c_1 \eta -  c_2 \eta^2 } \right],
        \end{align}
        for any $\eta <\frac{c_1}{c_2}$, where $c_1 = \frac{1-\epsilon_{\rm max} (1+\psi)}{2}$ and $c_2 = \beta(1+\epsilon_{\rm max})(1+\psi)$.
    \end{thm}
    \begin{IEEEproof}
        See Appendix~\ref{Apdx:Thm2}.
    \end{IEEEproof}
    \vspace{1mm}

    Theorem 2 shows that federated learning converges to a stationary point of a smooth loss function at the rate of $\mathcal{O}\big(\frac{1}{\sqrt{T}}\big)$, which is the same order as the original SGD algorithm \cite{SignSGD}, if the maximum normalized reconstruction error at the PS is lower than a certain level $\frac{1}{1+\psi}$. 
    Theorem~2 also shows that the convergence rate of federated learning improves as the reconstruction error of the global gradient vector at the PS reduces because  $c_1\eta - c_2\eta^2$ increases as $\epsilon_{\rm max}$ decreases. 
    Fortunately, the reconstruction error of the global gradient vector can be effectively reduced by employing the proposed gradient reconstruction algorithm, as will be demonstrated in Sec.~\ref{Sec:Simul}.
    Therefore, our analytical and numerical results together imply that  the convergence rate of federated learning can be improved using the proposed algorithm. 
    As discussed in Theorem 1, the reconstruction error of the proposed algorithm reduces as the compression ratio decreases; thereby, Theorem 2 also reveals the trade-off between the communication overhead and convergence rate of federated learning.

	\section{Simulation Results}\label{Sec:Simul}
	In this section, using simulation, we evaluate the performance of federated learning over a wireless MIMO MAC when employing the communication-efficient federated learning framework in Algorithm~\ref{alg:Framework}.
	We set $K=32$, $U=64$, and $\sigma^2=1$, while modeling the entries of the MIMO channel matrix as IID Gaussian random variables with zero mean and unit variance.
	We assume that the downlink communication is error free unless specified otherwise.
	For the gradient compression technique in Sec.~\ref{Sec:Comp}, we set $M = \lfloor N/R  \rfloor$ and $S = \lfloor S_{\rm ratio} N \rfloor$, where $S_{\rm ratio}$ is the ratio of the sparsification level.
	For the gradient reconstruction at the PS, we consider the following algorithms:

	\begin{itemize}
        \item {\bf Proposed algorithm:} This is given in Algorithm~\ref{alg:Proposed}.  
	    To initialize the algorithm, we set\footnote{In this initialization, we use $\mathbb{E} \big[|({\bf x}_{k,b}^{(t)})_m|^2\big] \approx {\|{\bf x}_k^{(t)}\|^2}/{\bar{M}} = {1}/{P_k^{(t)}}$ and $\mathbb{E} \big[|({\bf g}_{k,b}^{(t)})_n|^2\big] \approx \mathbb{E} \big[|({\bf x}_{k,b}^{(t)})_m|^2\big]/R \approx {1}/{(R P_k^{(t)})}$.} $\hat{x}_k^{\rm pri}[m] = 0$, $\nu_{x_k[m]}^{\rm pri} ={1}/{P_k^{(t)}}$, and ${\bm \nu}_{{\bf g}_{k,b}^{(t)}} = {1}/{(R P_k^{(t)})}{\bf 1}_{N}$.
	    An initial estimate $\hat{\bf g}_{k,b}^{(t)}$ is randomly drawn from $\mathcal{N}({\bf 0}_{N},{\bm \nu}_{{\bf g}_{k,b}^{(t)}})$.
	    The parameters ${\bm \theta}_{k,b}^{(t)}$ of the Bernoulli Gaussian-mixture model for ${\bf g}_{k,b}^{(t)}$ are initialized with $\lambda_0=0.9$, $L=3$, $\lambda_{\rm \ell} = \frac{1-\lambda_0}{L}$, $\mu_{\ell} = \hat{g}_{\rm min} + \frac{2\ell - 1}{2L} (\hat{g}_{\rm max} - \hat{g}_{\rm min})$, $\phi_{\ell} = \frac{1}{12}\left(\frac{\hat{g}_{\rm max} - \hat{g}_{\rm min}}{ L}\right)^2$, $\forall \ell\in\{1,\ldots,L\}$, where $\hat{g}_{\rm max}$ and $\hat{g}_{\rm min}$ are the largest and the smallest entry of $\hat{\bf g}_{k,b}^{(t)}$, respectively. 
	    Other parameters are set as $\tau_{\rm GAMP}=10^{-5}$, $I_{\rm GAMP}=30$, and $I_{\rm Turbo}=2$, unless otherwise specified.

	    \item {\bf LMMSE-OMP:} This is a variation of the proposed algorithm that employs the LMMSE MIMO detector followed by the OMP algorithm in \cite{Yonina:Book,CS:Book,Tropp:04} without harnessing the turbo decoding principle. 
	    In this algorithm, the distribution of the compressed gradient vector is modeled as ${\bf x}^{(t)}[m] \sim \mathcal{N}({\bf 0},{\bf P}^{(t)})$. 
	    The OMP algorithm is set to stop after $S$ iterations. 

	    \item {\bf 2D-OMP:} This is the OMP algorithm proposed in \cite{Yonina:13} that directly solves a matrix-form problem in \eqref{eq:received_signal_BL}.
	    This algorithm is set to stop after $S K$ iterations.

        \item {\bf Kron-OMP:} This is the OMP algorithm in \cite{Yonina:Book,CS:Book,Tropp:04} that solves a Kronecker-form problem in \eqref{eq:received_vector_Kron}. 
        This algorithm is set to stop after $S K$ iterations.

	\end{itemize}
	\noindent The computational complexities of the above algorithms are characterized in  Table~\ref{Table:Complex}.
	Note that in our simulations, different numbers of blocks, $B$, is adopted for different algorithms, in order to ensure that they have a similar complexity order, as specified in Table~\ref{Table:Complex}.
    Meanwhile, the same sub-vector index sets are adopted by every device (i.e., $\mathcal{I}_{k,b} = \mathcal{I}_{b}$, $\forall k$).

    \begin{table*}
    	\renewcommand{\arraystretch}{1.3}
    	\caption{The number of real multiplications required by gradient reconstruction algorithms.}\vspace{-3mm}
    	\label{Table:Complex}
    	\centering
    	\begin{tabular}{c||c|c|c}
    		\hline
    		\bfseries Algorithm & \makecell{Complexity order}
    		& \makecell{Value in simulation \\ for $R=3$, $S_{\rm ratio} = 4\%$}  
    		& \makecell{Value in simulation \\ for $R=5$, $S_{\rm ratio} = 4\%$}  
    		\\
    		\hline\hline
    		Proposed algorithm ($B=10$)
    		& $\left(U^3M + N M K I_{\rm GAMP} \right) I_{\rm Turbo}B$  
    		& $1.90 \times 10^{10}$ 
    		& $1.14 \times 10^{10}$ 
    		\\	
    		\hline
    		LMMSE-OMP ($B=10$)
    		& $\left(U^3M + KS^4/4  + N M KS \right) B$
    		& $1.97 \times 10^{10}$ 
    		& $1.23 \times 10^{10}$
    		\\	
    		\hline
    		2D-OMP ($B=100$)
    		& $\left\{(KS )^4/4 + (U+N)MK^2 S \right\}B$
    		& $3.67 \times 10^{10}$ 
    		& $3.82 \times 10^{10}$ 
    		\\
    		\hline
    		Kron-OMP ($B=300$)
    		& $\left\{(KS )^4/4 + UN M K^2 S \right\}B$
    		& $4.12 \times 10^{10}$ 
    		& $2.21 \times 10^{10}$ 
    		\\	
    		\hline
    	\end{tabular}
    	\vspace{-5mm}
    \end{table*}

	In this simulation, we consider an image classification task using the MNIST dataset in which  a handwritten digit (from $0$ to $9$) of a $28\times 28$ grayscale image is classified using a  neural network \cite{MNIST}.
	We particularly adopt the neural network that consists of $784$ input nodes, a single hidden layer with $20$ hidden nodes, and $10$ output nodes.  
	The activation functions of the hidden layer and the output layer are set as the ReLU and the softmax functions, respectively. 
	The weights of this neural network at iteration $t$ is mapped into a parameter vector ${\bf w}_t$ with $\bar{N}=15910$. 
	To train the network, we consider a mini-batch SGD algorithm with mini-batch size $|\mathcal{D}_k^{(t)}|=10$, a fixed learning rate $\eta_t  =0.2$, and the cross-entropy loss function.
	The local training data set of device~$k$, $\mathcal{D}_k$, is determined by randomly drawing $1000$ training data samples labeled with digit $d_k= \big\lfloor \frac{k-1}{K/10}\big\rfloor$ among $60000$ training data samples in the MNIST dataset; this corresponds to a \textit{non-IID} setting because each device has the information of only one digit.
	For performance evaluation, we consider two metrics: \textit{Classification accuracy} computed for $10000$ training data samples in the MNIST dataset, and \textit{normalized MSE (NMSE)} of the global gradient vector, defined as $\mathbb{E}[\epsilon_t]$ with $ \epsilon_t = \frac{\|{\bf g}_{\rm avg}^{(t)} - \hat{\bf g}_{\rm avg}^{(t)}\|^2}{\|{\bf g}_{\rm avg}^{(t)}\|^2}$.

	\begin{figure}
		\centering 
		\subfigure[Classification accuracy]
		{\epsfig{file=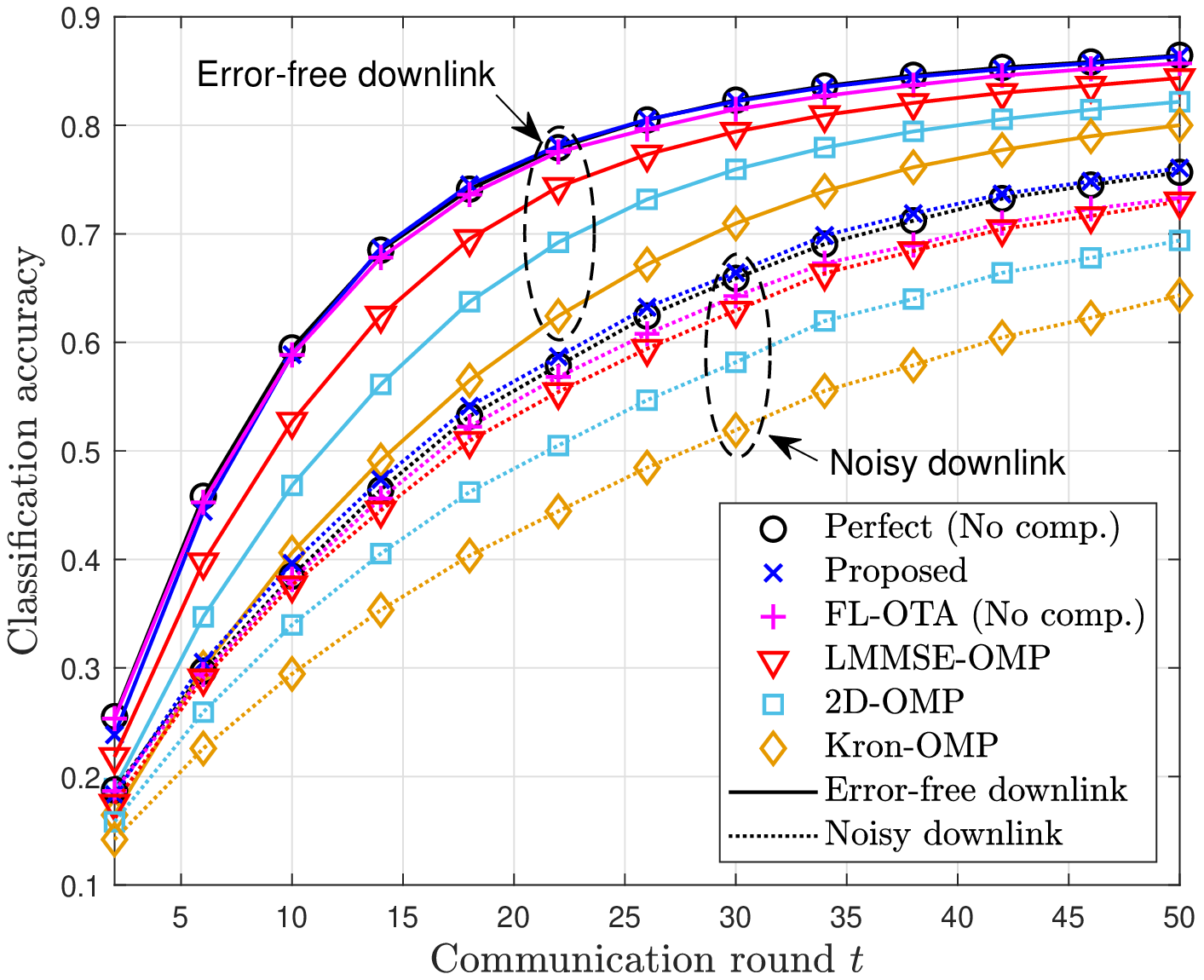, width=8.1cm}}
		\subfigure[Normalized MSE]
		{\epsfig{file=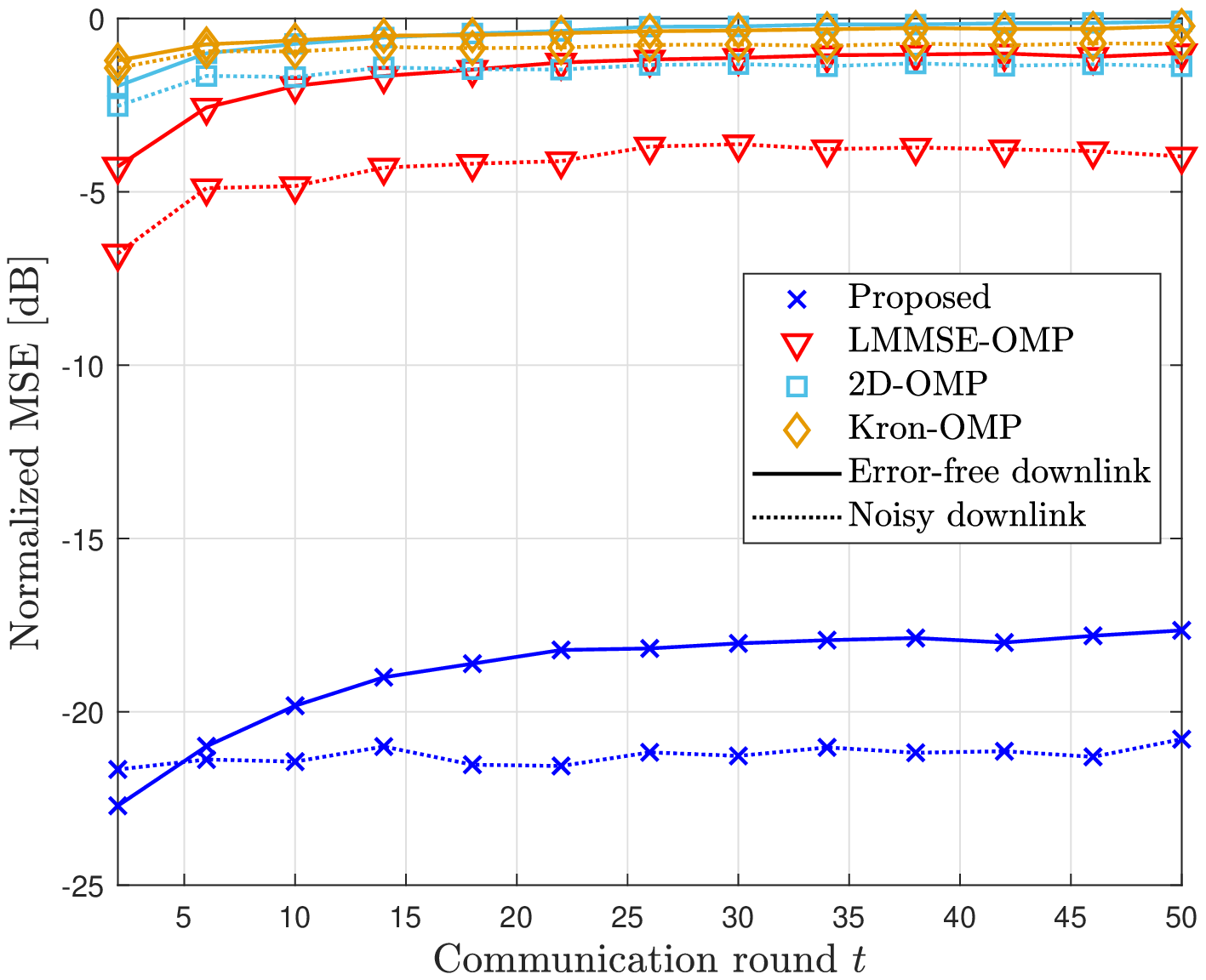, width=8.1cm}}
		\caption{Performance comparison of different gradient reconstruction algorithms when $R=5$ and $S_{\rm ratio} = 4 \%$.} \vspace{-5mm}
		\label{fig:Comp}
	\end{figure}

    In Fig.~\ref{fig:Comp}, we evaluate the classification accuracy and the NMSE of federated learning with different gradient reconstruction algorithms when $R=5$ and $S_{\rm ratio} = 4 \%$.
    In Fig.~\ref{fig:Comp}(a), we also plot the classification accuracy of federated learning based on over-the-air computation (FL-AirComp) developed in \cite{FL:IRS} by assuming no device selection and no intelligent reflecting surface between the PS and the wireless devices.
    In a noisy downlink communication scenario, we assume that the parameter vector received by device $k$ at iteration $t$ is modeled as ${\bf w}_{k}^{(t)} = \epsilon_{\rm DL} {\bf w}_t + \sqrt{1-\epsilon_{\rm DL}^2} {\bf e}_{k,t}$, where $\epsilon_{\rm DL} = 0.7$, ${\bf e}_{k,t}$ is an independent Gaussian random vector distributed as $\mathcal{N}\big({\bf 0}_{\bar{N}}, {\sf diag}(|w_{t,1}|^2,\ldots,|w_{t,\bar{N}}|^2)\big)$, and $w_{t,i}$ is the $i$-th entry of ${\bf w}_t$.
    Fig.~\ref{fig:Comp}(a) shows that for both the error-free and the noisy downlink scenarios, the proposed algorithm achieves almost the same classification accuracy as perfect reconstruction with no compression.
    Meanwhile, Fig.~\ref{fig:Comp}(b) shows that the proposed algorithm enables almost lossless (less than $-17$ dB) reconstruction of the global gradient vector at the PS.
    Unlike the proposed algorithm, conventional CS algorithms (2D-OMP and Kron-OMP) cause degradation in classification accuracy (see Fig.~\ref{fig:Comp}(a)) due to severe loss of reconstruction accuracy (see Fig.~\ref{fig:Comp}(b)).
    Although both the proposed algorithm and LMMSE-OMP adopt the divide-and-conquer strategy described in Sec.~\ref{Sec:Proposed}, the proposed algorithm outperforms LMMSE-OMP, attained by harnessing the turbo decoding principle with the EM-GAMP algorithm.
    FL-AirComp provides a similar classification accuracy as the proposed algorithm, but this framework assumes no gradient compression at the devices; thereby, AirComp requires  $R=5$ times higher communication overhead than the federated learning framework considered in the proposed algorithm.

	\begin{figure*}
    	\begin{minipage}[b]{8cm}
    		\centering
    	{\epsfig{file=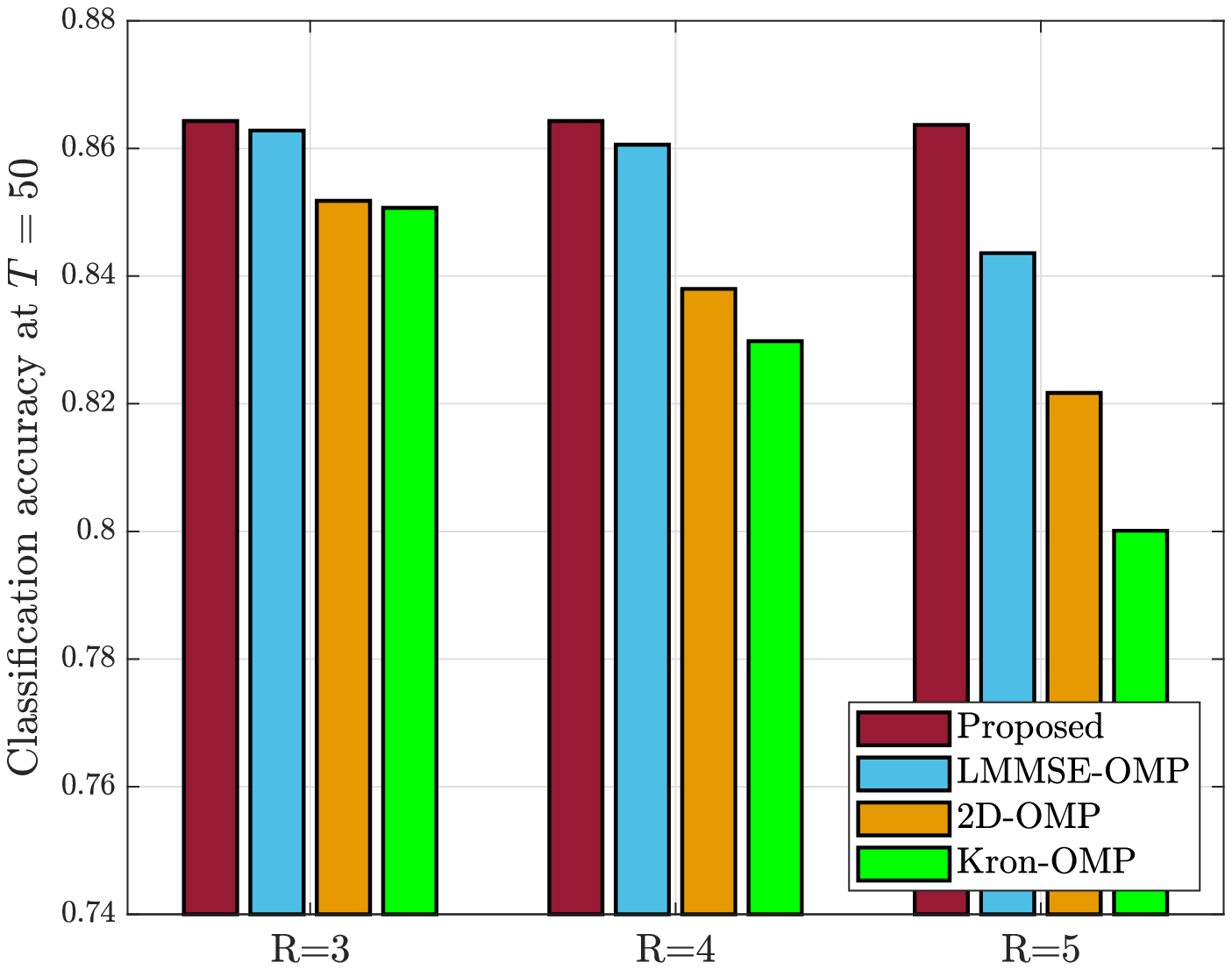, width=8cm}}\vspace{-5mm}
    	\caption{Classification accuracy of federated learning with different compression ratios when $S_{\rm ratio} = 4 \%$.} 
    	\label{fig:Ratio}
    	\end{minipage}\hfill
    	\begin{minipage}[b]{8cm}
    	\centering
    	{\epsfig{file=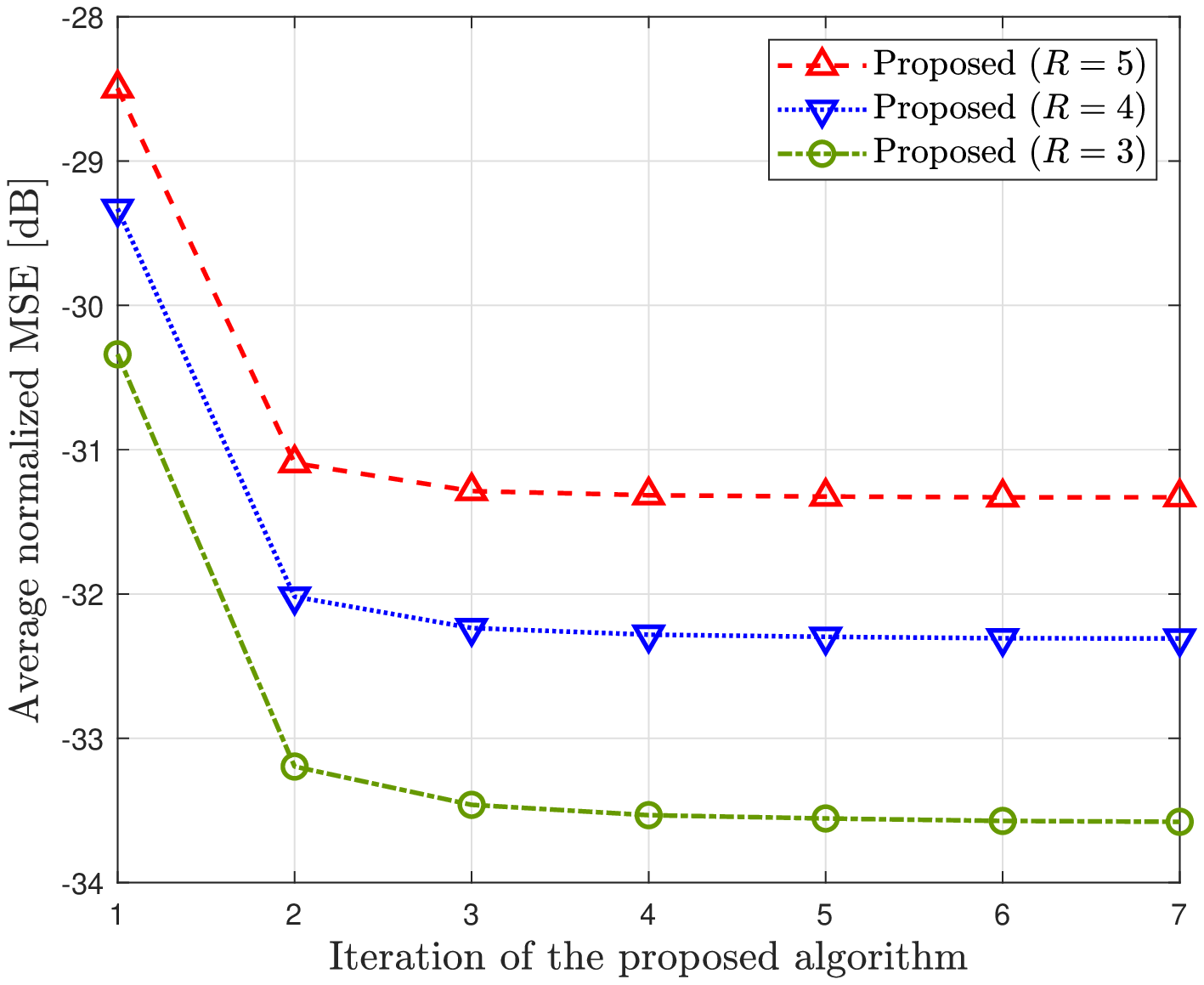, width=8cm}}\vspace{-5mm}
    	\caption{Average NMSE of the proposed algorithm with different numbers of turbo iterations when $S_{\rm ratio} = 1 \%$.} 
    	\label{fig:Turbo}
    	\end{minipage}
    	\vspace{-5mm}
    \end{figure*}

	In Fig.~\ref{fig:Ratio}, we evaluate the classification accuracy of federated learning with different compression ratios when $S_{\rm ratio} = 4 \%$.
	Fig.~\ref{fig:Ratio} shows that the classification accuracies of all the gradient reconstruction algorithms degrade with the compression ratio. 
	This result follows from the fact that the smaller the projected dimension in \eqref{eq:compress_block}, the more difficult the reconstruction of the local gradient vectors at the PS. 
	Since communication overhead of federated learning reduces with the compression ratio, this result also reveals the trade-off between communication overhead and learning performance of federated learning.  
	Despite this trade-off, the proposed algorithm provides almost the same accuracy even at the compression ratio of $R=5$, implying that the proposed algorithm has the best robustness against the increase in the compression ratio.  
	Therefore, the proposed algorithm minimizes the communication overhead of federated learning required to achieve a certain level of classification accuracy.

	\begin{figure}
		\centering 
		\subfigure[Average classification accuracy]
		{\epsfig{file=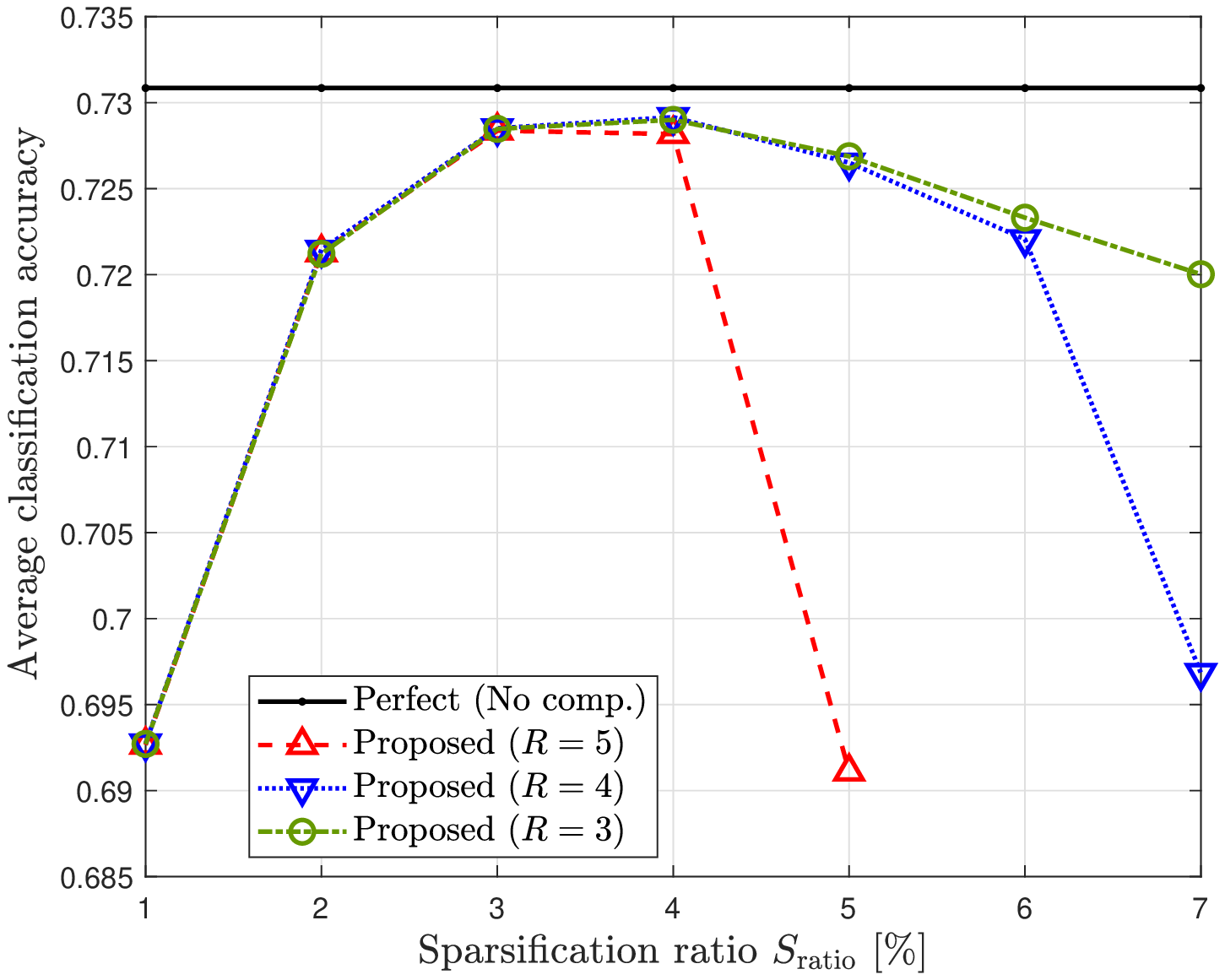, width=8.1cm}}
		\subfigure[Average NMSE]
		{\epsfig{file=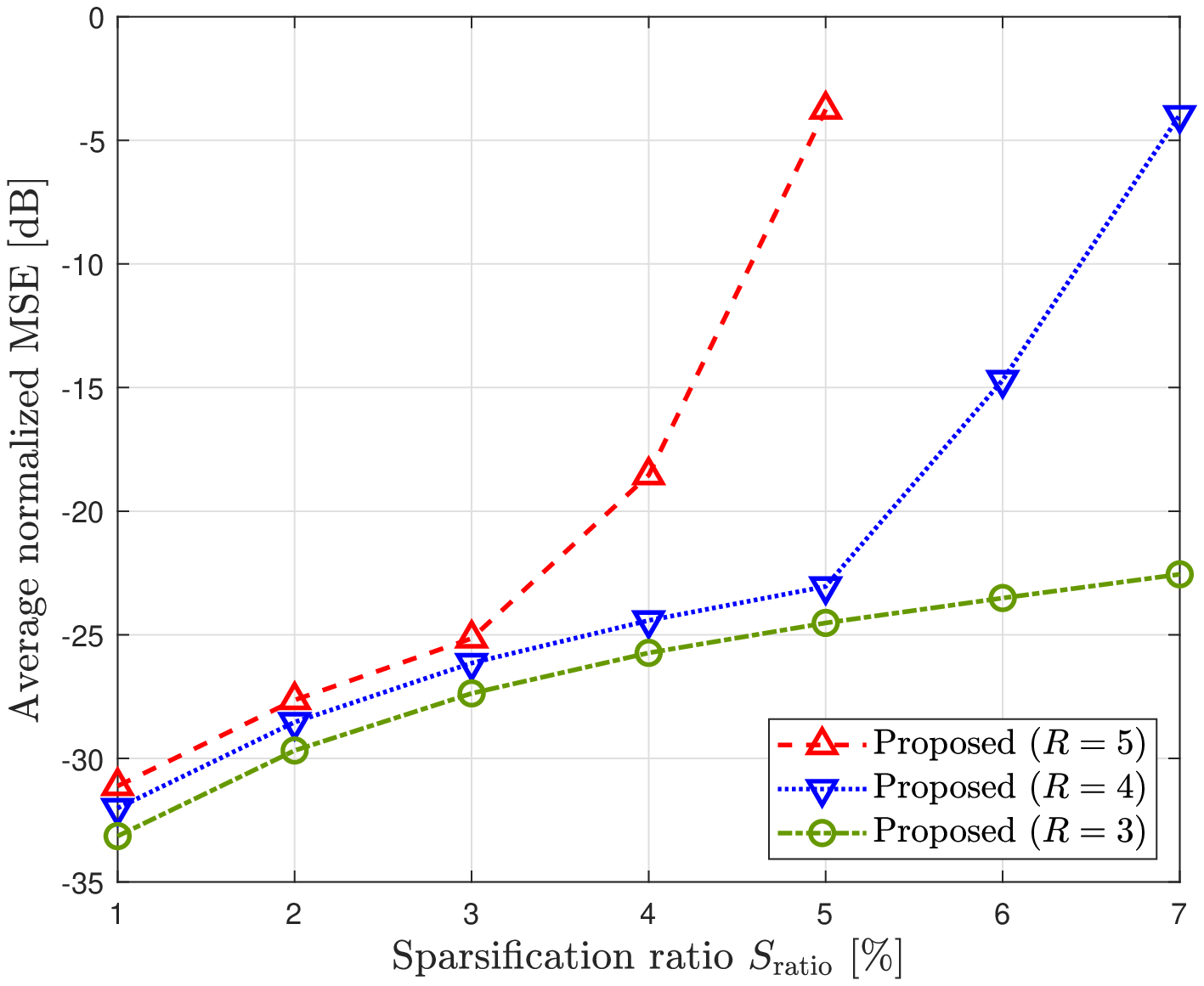, width=8.1cm}}
		\caption{Performance of the proposed gradient reconstruction algorithm with different sparsification ratios.} \vspace{-5mm}
		\label{fig:Sparse}
	\end{figure}

	In Fig.~\ref{fig:Turbo}, we evaluate the average NMSE of the proposed algorithm with different numbers of turbo iterations when $S_{\rm ratio} = 1 \%$.
	The NMSE is averaged over the first 20 iterations.
    Fig.~\ref{fig:Turbo} shows that the reconstruction accuracy of the proposed algorithm improves with the number of turbo iterations. 
    This performance improvement comes from the information exchange between Module A and Module B in the proposed algorithm; during this exchange, the information from the uplink received signals and the sparse property of the local gradient vector complement each other to improve the reconstruction accuracy of the local gradient vectors. 
    Since this effect vanishes as the information exchange repeats, the performance improvement of the proposed algorithm becomes marginal after a few iterations (i.e., 3 iterations).

    In Fig.~\ref{fig:Sparse}, we evaluate the average classification accuracy and NMSE of the proposed gradient reconstruction algorithm with different sparsification ratios.
    These performance measures are averaged over the first 50 iterations.
    Fig.~\ref{fig:Sparse}(a) shows that there exists the optimal sparsification ratio that provides the highest classification accuracy. 
    This result is due to the trade-off between the amount of the local gradient information conveyed to the PS and the reconstruction accuracy of the local gradient vectors at the PS. 
    To be more specific, increasing the sparsification ratio increases the number of non-zero gradient entries conveyed to the PS per iteration but, at the same time, decreases the sparsity of the local gradient vectors, which leads to the degradation in the reconstruction accuracy, as shown in Fig.~\ref{fig:Sparse}(b). 
    Our simulation results imply that the performance of the proposed algorithm can be maximized by optimizing the selection of the sparsification level, which would be interesting for future work.

	\section{Conclusion}
	In this paper, we have studied communication-efficient federated learning over a wireless MIMO MAC in which communication overhead for local gradient transmission is effectively reduced by employing gradient compression at wireless devices and simultaneous transmission via shared radio resources. 
    Under this scenario, we have presented a turbo-iterative algorithm that solves a computationally demanding gradient reconstruction problem at the PS based on a divide-and-conquer strategy.
	One prominent feature of the presented algorithm is that gradient reconstruction accuracy at the PS improves as the iteration continues. In each iteration, the belief from uplink received signals at the PS is complemented by the sparse property of the local gradient vectors. 
    We have also characterized the reconstruction error of the presented algorithm and a sufficient condition to guarantee the convergence of federated learning.  
	Using simulations, we have demonstrated that the presented algorithm achieves almost the same performance as a perfect reconstruction scenario while providing a superior performance compared to existing CS algorithms.
	
	An important direction for future research is to optimize device scheduling and power control for uplink communication in federated learning, taking into account the different locations of devices in wireless networks. 
	It would also be important to study downlink communication of federated learning to enable accurate and efficient broadcasting of the global model to wireless devices. 

	\appendices
	\section{Proof of Lemma~1}\label{Apdx:Lem1}	
	Suppose that the $b$-th local gradient sub-vector at device $k$, ${\bf g}_{k,b}^{(t)}$, follows an IID random distribution with mean $\mu_{g_{k,b}}^{(t)}$ and variance $\zeta_{k,b}^{(t)}$.
	Since ${\bf A}^{(t)}$ is an IID random matrix with $({\bf A}^{(t)})_{m,n}\sim \mathcal{N}(0,1/M)$, the projection in \eqref{eq:compress_block} implies that ${\bf x}_{k,b}^{(t)}\sim\mathcal{N}({\bf 0}_M,R \zeta_{k,b}^{(t)} {\bf I}_M)$ for large $N$ by the central limit theorem. 
    Then from \eqref{eq:x_hat_post} and \eqref{eq:R_x_post}, the posterior mean and variance of $x_k^{(t)}[m]$ for $m\in\mathcal{M}_b$ are computed as 
	\begin{align}
        \hat{x}_k^{\rm post}[m] 
    	&=   \nu_{x_k[m]}^{\rm post} \left\{ 
    	\sum_{j=1}^K \chi_{k,b,j}^{(t)}  x_j^{(t)}[m] + \chi_{k,b,{\bf z}}^{(t)}[m]  \right\},~~\text{and}~~ \nu_{x_k[m]}^{\rm post}
    	= \frac{R\zeta_{k,b}^{(t)}}{1 + R \zeta_{k,b}^{(t)}\chi_{k,b,k}^{(t)} },  \label{eq:x_hat_post3} 
	\end{align}	
	respectively, where $\chi_{k,b,j}^{(t)} = P_k^{(t)} \big({\bf h}_k^{(t)}\big)^{\sf T} \big({\bm \Phi}_{k,b}^{(t)} + \sigma^2 {\bf I}_U\big)^{-1} {\bf h}_j^{(t)}$, 
	$\chi_{k,b,{\bf z}}^{(t)}[m] = \sqrt{P_k^{(t)}}\big({\bf h}_k^{(t)}\big)^{\sf T} \big({\bm \Phi}_{k,b}^{(t)} + \sigma^2 {\bf I}_U\big)^{-1}{\bf z}^{(t)}[m]$, ${\bm \Phi}_{k,b}^{(t)} = R \sum_{j\neq k}  \zeta_{j,b}^{(t)}P_j^{(t)}{\bf h}_j^{(t)}\big({\bf h}_j^{(t)}\big)^{\sf T}$,
	and ${\bf h}_k^{(t)}$ is the $k$-th column of ${\bf H}^{(t)}$.
    By plugging $\hat{x}_k^{\rm post}[m]$, $\nu_{x_k[m]}^{\rm post}$, $\hat{x}_k^{\rm pri}[m] = 0$, and $\nu_{x_k[m]}^{\rm pri}=R \zeta_{k,b}^{(t)}$ into \eqref{eq:x_hat_nu_ext}, the extrinsic mean and variance of $x_k^{(t)}[m]$ for $m\in\mathcal{M}_b$ are computed as 
    \begin{align}
    	\hat{x}_k^{(t)}[m]  =  
    	x_k^{(t)}[m] + \frac{1}{\chi_{k,b,k}^{(t)}} \Bigg\{\sum_{j\neq k} \chi_{k,b,j}^{(t)} x_j^{(t)}[m]  + \chi_{k,b,{\bf z}}^{(t)}[m] \Bigg\},
    	~~\text{and}~~
    	{\nu}_{k,b}^{(t)}
    	= \frac{1}{\chi_{k,b,k}^{(t)}}, \label{eq:ext_x_kt} 
    \end{align}	
    respectively.
    From the eigendecomposition of ${\bm \Phi}_{k,b}^{(t)}$, the extrinsic variance  $\nu_{k,b}^{(t)}$ can be rewritten as
    \begin{align}\label{eq:nu_express}
        \nu_{k,b}^{(t)}
        &= \frac{1}{{\sf SNR}_k^{(t)}}  \Bigg( \|{\bf h}_k^{(t)}\|^2 -  \sum_{i=1}^{E_{k,b}^{(t)}}  \big|\big({\bf u}_{k,b,i}^{(t)}\big)^{\sf T} {\bf h}_k^{(t)}\big|^2  \frac{\lambda_{k,b,i}^{(t)}}{\lambda_{k,b,i}^{(t)} + \sigma^2}  \Bigg)^{-1},
    \end{align}
    where ${\sf SNR}_k^{(t)} = \frac{P_k}{\sigma^2}$, $\lambda_{k,b,i}^{(t)}$ is the $i$-th largest eigenvector of ${\bm \Phi}_{k,b}^{(t)}$, ${\bf u}_{k,b,i}^{(t)}$ is the eigenvector associated with $\lambda_{k,b,i}^{(t)}$, and $E_{k,b}^{(t)}$ is the rank of ${\bm \Phi}_{k,b}^{(t)}$.
	Since ${\bf x}_{k,b}^{(t)}$ is an IID Gaussian vector for large $N$, the extrinsic mean $\hat{\bf x}_{k,b}^{(t)} = \big[\hat{x}_k^{(t)}[\mathcal{M}_b(1)],\cdots, \hat{x}_k^{(t)}[\mathcal{M}_b(M)] \big]^{\sf T}$ becomes an AWGN observation of ${\bf x}_{k,b}^{(t)}$ where the noise variance is the extrinsic variance ${\nu}_{k,b}^{(t)}$ \cite{Liu:19}. This completes the proof.

	\section{Proof of Theorem~\ref{Thm1}}\label{Apdx:Thm1}	
	Suppose that the $b$-th local gradient sub-vector at device $k$, ${\bf g}_{k,b}^{(t)}$, follows an IID random distribution with mean $\mu_{g_{k,b}}^{(t)}$ and variance $\zeta_{k,b}^{(t)}$.
	Under the assumption that the distributions of $\{{\bf g}_{k,b}^{(t)}\}_k$ are known at the PS in prior, the EM-GAMP algorithm in Module B of the proposed algorithm reduces to the GAMP algorithm in \cite{Rangan:ISIT} which does not have the EM step (i.e., without Step 8 in Algorithm~\ref{alg:EMGAMP}). 
	In the asymptotic regime of $N \rightarrow \infty$ and $N/M \rightarrow R\geq 1$, the behavior of the GAMP algorithm applied to reconstruct ${\bf g}_{k,b}^{(t)}$ from $\hat{\bf x}_{k,b}^{(t)}$ is characterized by a scalar state evolution.
    If this state evolution has a unique fixed point, the output of the GAMP algorithm, $\hat{\bf g}_{k,b}^{(t)}$, converges to the MMSE estimate of ${\bf g}_{k,b}^{(t)}$ for a given observation $\hat{\bf x}_{k,b}^{(t)}$ \cite{Guo:TIT,Bayati:TIT,Rangan:ISIT}.	
    Therefore, the MSE between ${\bf g}_{k,b}^{(t)}$ and $\hat{\bf g}_{k,b}^{(t)}$ is less than or equal to the MSE between ${\bf g}_{k,b}^{(t)}$ and its linear MMSE estimate, i.e.,
    \begin{align}
        \mathbb{E}\big[\|  {\bf g}_{k,b}^{(t)} - \hat{\bf g}_{k,b}^{(t)} \|^2\big] 
        &\leq \mathbb{E}\big[\| {\bf g}_{k,b}^{(t)} - \hat{\bf g}_{{\rm LMMSE},k,b}^{(t)}\|^2\big], 
        \label{eq:LMMSE_bound}
    \end{align}    
    where $\hat{\bf g}_{{\rm LMMSE},k,b}^{(t)}$ is the linear MMSE estimate of ${\bf g}_{k,b}^{(t)}$ for a given observation $\hat{\bf x}_{k,b}^{(t)}$ in \eqref{eq:Lem1}.
    From \eqref{eq:Lem1} with ${\bf x}_{k,b}^{(t)} = {\bf A}^{(t)}{\bf g}_{k,b}^{(t)}$, the MSE of the linear MMSE estimate is computed as \cite{Kay:Book}
    \begin{align}
        \mathbb{E}\big[\| {\bf g}_{k,b}^{(t)} - \hat{\bf g}_{{\rm LMMSE},k,b}^{(t)}\|^2\big] 
        &=  N\zeta_{k,b}^{(t)}  -  \big(\zeta_{k,b}^{(t)}\big)^2 {\sf Tr}\left[ ({\bf A}^{(t)})^{\sf T} \left( \zeta_{k,b}^{(t)} {\bf A}^{(t)} ({\bf A}^{(t)})^{\sf T} +  \nu_{k,b}^{(t)}{\bf I}_{M}\right)^{-1} {\bf A}^{(t)}  \right] 
        \nonumber \\
        &\overset{(a)}{=}
        N \zeta_{k,b}^{(t)} \left[1 - \frac{ \zeta_{k,b}^{(t)}}{R\zeta_{k,b}^{(t)} + \nu_{k,b}^{(t)} } \right],
        \label{eq:LMMSE_MSE}
    \end{align}    
    where the equality (a) follows from ${\bf A}^{(t)}({\bf A}^{(t)})^{\sf T} \rightarrow R{\bf I}_M$ as $N \rightarrow \infty$ under the assumption that ${\bf A}^{(t)}$ is an IID random matrix with $({\bf A}^{(t)})_{m,n}\sim \mathcal{N}(0,1/M)$.
    Since the $k$-th local gradient vector is reconstructed as $\hat{\bf g}_{k}^{(t)} = {\sf Concatenate}\big(\{\hat{\bf g}_{k,b}^{(t)}\}_{b=1}^B\big)$,
    the MSE between $\hat{\bf g}_{\rm avg}^{(t)} = \sum_{k=1}^K \rho_k \hat{\bf g}_{k}^{(t)}$ and ${\bf g}_{\rm avg}^{(t)} = \sum_{k=1}^K \rho_k{\bf g}_{k}^{(t)}$ is expressed as
    \begin{align}\label{eq:global_MSE}
        \mathbb{E}\big[\| {\bf g}_{\rm avg}^{(t)} - \hat{\bf g}_{\rm avg}^{(t)}\|^2\big]
        =\mathbb{E}\left[\left\| \sum_{k=1}^K \rho_k  \big( {\bf g}_k^{(t)} - \hat{\bf g}_{k}^{(t)} \big)\right\|^2\right] 
        \overset{(a)}{=}\sum_{k=1}^K \rho_k^2 \sum_{b=1}^B \mathbb{E}\big[\|  {\bf g}_{ k,b}^{(t)} - \hat{\bf g}_{k,b}^{(t)} \|^2\big], 
    \end{align}
    where the equality (a) holds because different parameters in the global model are simultaneously transmitted by different devices when $\mathcal{I}_{k,b} \neq \mathcal{I}_{j,b}$ for $k\neq j$.
    Applying \eqref{eq:LMMSE_bound} and \eqref{eq:LMMSE_MSE} into \eqref{eq:global_MSE} yields the result in \eqref{eq:Thm1}.


	\section{Proof of Theorem~\ref{Thm2}}\label{Apdx:Thm2}	 
    Under Assumption 1, the improvement of the loss function at iteration $t$ is upper bounded as
    \begin{align}
        &F({\bf w}_{t+1}) - F({\bf w}_t)  
        \leq  \nabla F({\bf w}_t)^{\sf T}({\bf w}_{t+1}-{\bf w}_t)
            +  \frac{\beta}{2}\|{\bf w}_{t+1}-{\bf w}_t\|^2 \nonumber \\
        &\overset{(a)}{=} 
        - \eta_t \nabla F({\bf w}_t)^{\sf T} {\bf g}_{\rm avg}^{(t)} 
        - \eta_t \nabla F({\bf w}_t)^{\sf T} {\bf e}_{\rm avg}^{(t)} 
            +  \eta_t^2\frac{\beta}{2}\| {\bf g}_{\rm avg}^{(t)} +  {\bf e}_{\rm avg}^{(t)}  \|^2 \nonumber\\
        &\overset{(b)}{\leq}  - \eta_t  \nabla F({\bf w}_t)^{\sf T} {\bf g}_{\rm avg}^{(t)}   + \frac{\eta_t}{2} \big( \|\nabla F({\bf w}_t)\|^2 + \|{\bf e}_{\rm avg}^{(t)} \|^2 \big) 
        + \eta_t^2\beta \big( \|{\bf g}_{\rm avg}^{(t)} \|^2 + \|{\bf e}_{\rm avg}^{(t)}  \|^2 \big),
        \label{eq:loss_improve0}
    \end{align}
    where the equality (a) follows from \eqref{eq:global_update} with ${\bf e}_{\rm avg}^{(t)} =  \hat{\bf g}_{\rm avg}^{(t)} - {\bf g}_{\rm avg}^{(t)}$, and the inequality (b) follows from $\pm {\bf a}^{\sf T}{\bf b} \leq  \frac{1}{2}(\|{\bf a}\|^2 + \|{\bf b}\|^2)$.
    Let $\epsilon_t = { \| \hat{\bf g}_{\rm avg}^{(t)} - {\bf g}_{\rm avg}^{(t)} \|^2}/{\|{\bf g}_{\rm avg}^{(t)} \|^2}$ be a normalized reconstruction error of the global gradient vector at iteration $t$.
    Then the inequality in \eqref{eq:loss_improve0} is rewritten as
    \begin{align}
        F({\bf w}_{t+1}) - F({\bf w}_t)  
        &\leq  - \eta_t  \nabla F({\bf w}_t)^{\sf T} {\bf g}_{\rm avg}^{(t)}   +\frac{\eta_t}{2} \big( \|\nabla F({\bf w}_t)\|^2 + \epsilon_t\|{\bf g}_{\rm avg}^{(t)}\|^2 \big) 
        + \eta_t^2\beta (1 +\epsilon_t ) \|{\bf g}_{\rm avg}^{(t)} \|^2,  
        \label{eq:loss_improve}
    \end{align}    
    Since $\mathbb{E} \big[{\bf g}_{\rm avg}^{(t)} \big| {\bf w}_t \big] = \nabla F({\bf w}_t)$ and $\mathbb{E} \big[\|{\bf g}_{\rm avg}^{(t)} \|^2 \big| {\bf w}_t \big] \leq (1+\psi)  \|\nabla F({\bf w}_t) \|^2$ under Assumption 2, taking the expectation of both sides of \eqref{eq:loss_improve} conditioned on ${\bf w}_t$ yields
    \begin{align}\label{eq:improve_bound}
        \mathbb{E}\big[ F({\bf w}_{t+1}) - F({\bf w}_t) \big| {\bf w}_t\big]  
        &\leq  - \frac{\eta_t}{2} \big\{ 1- \epsilon_t (1+\psi)\big\} \|\nabla F({\bf w}_t)\|^2
        + \eta_t^2\beta (1 +\epsilon_t ) (1+\psi) \|\nabla F({\bf w}_t)\|^2 \nonumber \\
        & = -(c_1 \eta_t  - c_2 \eta_t^2) \|\nabla F({\bf w}_t)\|^2,
    \end{align}     
    where $c_1 = \frac{1-\epsilon_{\rm max}(1+\psi)}{2}$, $c_2=\beta (1+\epsilon_{\rm max}) (1+\psi) > 0$, and $\epsilon_{\rm max} = \max_{t} \epsilon_t$.

    A lower bound of the initial loss with ${\bf w}_1$ is expressed as 
    \begin{align}
        F({\bf w}_{1}) - F({\bf w}^\star) 
        &\geq F({\bf w}_{1})  - \mathbb{E}[F({\bf w}_{T+1})]
        = \sum_{t=1}^{T} \mathbb{E}\left[ F({\bf w}_{t}) - F({\bf w}_{t+1})  \right] \nonumber \\
        &\overset{(c)}{\geq} \mathbb{E}\left[ \sum_{t=1}^{T} \left( c_1 \eta_t -  c_2 \eta_t^2 \right) \|\nabla F({\bf w}_t)\|^2 \right],
        \label{eq:loss_bound}
    \end{align}    
    where the inequality (c) follows from \eqref{eq:improve_bound}.
    Plugging an adaptive learning rate $\eta_t = \frac{\eta}{\sqrt{t}}>0$ into \eqref{eq:loss_bound} yields \cite{Lee:21} 
    \begin{align}\label{eq:loss_bound2}
        F({\bf w}_{1}) - F({\bf w}^\star) 
        &\geq
        \mathbb{E}\left[\sum_{t=1}^{T} \left( \frac{c_1 \eta}{\sqrt{t}} -  \frac{c_2 \eta^2 }{t} \right)  \|\nabla F({\bf w}_t)\|^2 \right].
    \end{align}
    If $c_1>0$, from $\frac{1}{t}\leq \frac{1}{\sqrt{t}}$ and $\frac{1}{\sqrt{T}} \leq \frac{1}{\sqrt{t}}$ for $1\leq t\leq T$, the right-hand-side of \eqref{eq:loss_bound2} is further lower bounded as  
    \begin{align}
        F({\bf w}_{1}) - F({\bf w}^\star) 
        &\geq
        \left( c_1 \eta -  c_2 \eta^2 \right) \mathbb{E}\left[ \frac{1}{\sqrt{T}}  \sum_{t=1}^{T}  \|\nabla F({\bf w}_t)\|^2 \right].
        \label{eq:loss_bound3}
    \end{align}
    Since $c_1>0$ for $\epsilon_{\rm max} < \frac{1}{1+\psi}$, the inequality in \eqref{eq:loss_bound3} is rewritten as in \eqref{eq:Thm2} for any $\eta < \frac{c_1}{c_2}$ with $\epsilon_{\rm max} < \frac{1}{1+\psi}$. 

\end{document}